\documentclass[11pt,twoside]{report}
\usepackage{a4wide,caption,epsfig,fancyhdr,url}
\usepackage{verbatim}
\usepackage{float}
\usepackage{graphicx}
\usepackage{amsmath}
\usepackage{tabularx}
\usepackage{multicol}
\usepackage{subcaption}
\usepackage{algorithm}
\usepackage{algpseudocode}
\usepackage{array}
\usepackage{multirow}
\usepackage{hyperref}
\usepackage{adjustbox}
\usepackage[margin=25mm]{geometry}
\usepackage{tabularray-2021}

\usepackage{mwe}
\usepackage{longtable} 

\usepackage{cellspace}
\usepackage{setspace}

\usepackage{multirow}

\newcommand{\probP}{\text{I\kern-0.15em P}} 
\date{August 08, 2022}
\newcommand{\SubmissionDate}{August 08, 2022}
\newcommand{\student}{Name: Mesfer Mohammed Alqarni}
\newcommand{\supervisor}{Supervisor: Ron Van Schyndel}
\newcommand{\topic}{A Framework to Allow a Third Party to Watermark Numerical Data in an Encrypted Domain while Preserving its Statistical Properties}
\newcommand{\school}{School of Computing Technologies}
\newcommand{\program}{Masters of Computer Science}
\newcommand{\collage} {STEM College}
\newcommand{\institution}{Royal Melbourne Institute of Technology University}

\begin{document}
\onehalfspacing 

\title{{\Large\bf \topic}}
\author{
A minor thesis submitted in partial fulfilment of the requirements for the degree of
\\\program\\*[10mm]
\\\student
\\\school
\\\collage
\\\institution
\\Melbourne, Victoria, Australia}

\maketitle
\thispagestyle{empty}

\chapter*{Declaration}

I declare that this thesis contains my own original work except where acknowledged. A part of this thesis which was also my own work, was previously submitted to RMIT University.

This work commenced in July 2021, under the
supervision of {\supervisor}.

\paragraph{}
\vspace{5cm}\noindent \\\student \\
\school\\
\collage\\
\institution\\
\SubmissionDate

\pagenumbering{roman}

\chapter*{Acknowledgements}

First and foremost, I would like to praise Allah the Almighty, the Most Gracious, and the Most Merciful for His blessing given to me during my study and in completing this thesis. Secondly, I would like to express my most profound appreciation to my supervisor Dr Ron Van Schyndel, for his guidance through the process of conducting this research. Also, I cannot begin to express my thanks to Wen-jie Lu from Alibaba Group, who provided endless support in terms of homomorphic encryption. Also, I would like to extend my thanks to my sponsor the Saudi Arabian Cultural Mission (SACM), for their support during my studies. Finally, I would like to thank my family, and friends for providing endless support.

\tableofcontents
\listoffigures
\listoftables

\newpage
\begin{table}[H]
\caption{The meanings of abbreviations and acronyms used in the thesis}
\begin{tabular}{ll}
\hline
\multicolumn{1}{c}{\textbf{Abbreviation}} & \multicolumn{1}{c}{\textbf{Meaning}}                                                        \\ \hline
2D                                          & Two-dimensional                                                                           \\
3D                                          & Three-dimensional                                                                         \\
AES                                         & Advanced encryption standard                                                              \\
BFV                                         & Brakerski/Fan-Vercauteren                                                                 \\
BGN                                         & Boneh-Goh-Nissim                                                                          \\
BGV                                         & Brakerski, Gentry, and Vaikuntanathan                                                     \\
CDW                                         & Content dependent watermarking                                                            \\
CKKS                                        & Cheon, Kim, Kim and Song (scheme)                                                           \\
CSV                                         & Comma-separated values                                                                    \\
DO                                          & Data obfuscation                                                                          \\
DW                                          & Digital watermarking                                                                      \\
DWT                                         & Discrete wavelet transform                                                                \\
FHEW                                        & Fastest homomorphic encryption in the west                                                \\
GB                                          & Gigabyte                                                                                  \\
GHz                                         & Gigahertz                                                                                 \\
HE                                          & Homomorphic encryption                                                                    \\
IoT                                         & Internet of Things                                                                        \\
KW                                          & Kilowatt                                                                                  \\ 
LUT                                         & Look-Up Table                                                                             \\
LWE                                         & Learning with errors                                                                      \\
MB                                          & Megabyte                                                                                  \\
ML                                          & Machine learning                                                                          \\
MRA                                         & Multi-resolution analysis                                                                 \\
ms                                          & Milliseconds                                                                              \\ 
N/A                                         & Not applicable                                                                            \\
OPE                                         & Order-preserving encryption                                                               \\
PRNG                                        & Pseudorandom number generator                                                             \\
RLWE                                        & Ring learning-with-errors                                                                 \\
SEAL                                        & Simple encrypted arithmetic library                                                       \\
SIMD                                        & Single instruction, multiple data                                                          \\
TFHE                                        & Fast fully homomorphic encryption scheme over the torus                                   \\
WHT                                         & Walsh-Hadamard transform                                                                  \\\hline
\end{tabular}
\end{table}

\newpage

\begin{longtable}[c]{S{m{0.1\linewidth}} S{m{0.83\linewidth}}}
\caption{The definitions of symbols used in the thesis}
       \\\hline   
        \textbf{Symbol} & \textbf{Definition}
            \\\hline \endfirsthead
            \hline
    $X$                                     & An unwatermarked dataset                                                              \\
    $X^{\prime}$                            & A watermarked dataset                                                                 \\
    $\overline{X}$                          & The mean of dataset $X$                                                           \\
    $\overline{X^{\prime}}$                 & The mean of dataset $X^{\prime}$                                                  \\
    $G$                                     & Pseudorandom number generator                                                         \\
    $K$                                     & The watermark                                                                         \\
    $n$                                     & The length of the dataset, and the watermark sequence                                 \\
    $SH_{key}$                              & A secret key used to shuffle the data                                                 \\
    $S$                                     & The sequence generated via the PRNG seeded with $K$                                   \\
    $\hat{X}$                               & A copy of the data expected to be watermarked                                         \\
    $S_{X}^{2}$                             & Variance of the dataset $X$                                                          \\
    $S_{X^{\prime}}^{2}$                    & Variance of the dataset $X^{\prime}$                                                 \\
    $s_{i}$                                 & Individual element of sequence $S$ at position $i$                            \\
    $P$                                     & The probability of a variable                                                        \\
    $T$                                     & Sequence follows a Gaussian $N(0,1)$ distribution                                     \\
    $t_{i}$                                 & Individual element of sequence $T$ at position $i$                            \\
    $M$                                     & A secret value kept by the watermarker for watermarking algorithm 2              \\
    $\epsilon$                              & Determines the error in the results of the watermarked dataset $X^{\prime}$           \\
    a, b, $\lambda$                         & Parameters used to watermark the dataset $X$ using watermarking algorithm 2       \\
    $x_{i}$                                 & Individual element of sequence $X$ at position $i$                            \\
    $x_{i}^{\prime}$                        & Individual element of sequence $X'$ at position $i$                           \\
    $\left|t_{i}\right|$                    & Absolute value of $t_{i}$                                                             \\
    $\hat{M}$                               & The verified secret value in the watermarked data $\hat{X}$                           \\
    $E(.)$                                  & Encryption function                                                                   \\
    $\mathcal{O}$                           & Represents the time complexity                                                                       \\
    $\overline{n}$                          & Represents the RLWE dimension                                                         \\
    $q$                                     & The ciphertext modulus                                                                \\
    $X^{\prime\prime}$                      & Watermarked and obfuscated dataset                                                    \\
    $Obf_{key}$                             & Secret key used to obfuscate and deobfuscate data                                     \\
    $\hat{x}_{i}$                           & Individual element of sequence $\hat{X}$ at position $i$                      \\
    $C$                                     & A column of data                                                                      \\
    $r$                                     & The number of ones in a binary sequence                                               \\
    $Bob_{RK}$                              & Bob's repacking key                                                                   \\
    $Bob_{SwK}$                             & Bob's key for the key-switch function                                                 \\
    $Bob_{pk}$                              & Bob's public key                                                                      \\
    $Bob_{EK}$                              & Bob's evaluation key                                                                  \\
    $Bob_{RotK}$                            & Bob's rotation key                                                                    \\
    $Bob_{RelK}$                            & Bob's relinearisation key                                                             \\
    $Bob_{s}$                               & Bob's secret key                                                                      \\
    $\textbf{g}_{digit}$                    & The digit decomposition gadget                                                        \\
    $\textbf{g}_{rns}$                      & The RNS decomposition gadget                                                          \\
    $\Delta$                                & The scaling factor                                                                    \\
    $\underline{n}$                         & Represents the LWE dimension for the input into the look-up table                     \\
    $n^\prime$                              & Represents the LWE dimension for the output of the look-up table                      \\
    
    $B_{ks}$                                & The digit decomposition base                                                          \\
    $h$                                     & The hamming weight                                                                    \\
    $msg$                                   & The message interval that the look-up table can process                     \\
    $S2C$                                   & Slot to coefficient function                                                          \\
    $moduli$                                & Represents the multiplicative depth for a ciphertext                                  \\
    $\mu$s                                  & Microseconds                                                                          \\\hline
\end{longtable}

\setlength{\parindent}{20pt} 

\chapter*{Summary}
This research contributed by building a framework that allows a neutral user to insert a specific mark (watermark), for source tracking purposes into secured data without the need to reveal the actual content of the data. In case the data is leaked, the neutral user will take on the role of judge and, using the inserted watermark, help the authorities to track and identify the user who leaked the data. The main contribution of this research is to increase the robustness of the inserted watermark in a way whereby other users cannot verify the watermark even if they know its actual value; consequently, they cannot remove the watermark either. The proposed method is efficient, does not affect the data usability for selected statistical operations, and is secure enough to protect the watermark from being removed.

\chapter*{Abstract}
\paragraph*{}
Watermarking data for source tracking applications by its owner can be unfair for recipients because the data owner may redistribute the same watermarked data to many users. Hence, each data recipient should know the watermark embedded in their data; however, this may enable them to remove it, which violates the watermark security. To overcome this problem, this research develops a framework that allows the cloud to watermark numerical data taking into consideration: the correctness of the results of selected statistics, data privacy, the recipient’s right to know the watermark that is embedded in their data, and the security of the watermark against passive attacks. The proposed framework contains two irreversible watermarking algorithms, each can preserve the correctness of the results for certain statistical operations. To preserve data privacy, the framework allows the cloud to watermark data while it is encrypted. Furthermore, the framework robustifies the security of the chosen algorithms to nominate the cloud as the only neutral judge able to verify the data ownership even if other users know the watermark. The security is enhanced in a way that does not affect the data usability. The time complexity to find the watermark is $\mathcal{O}(\frac{n !}{r !(n-r)!})$.

\newpage
\setcounter{page}{1}
\pagenumbering{arabic}

\chapter{Introduction}

    \section{Introduction}
    \paragraph{}
    Nowadays, the need to share data with other users is more important than ever, especially with the widespread use of Internet of Things (IoT) devices. The IoT is a set of physical devices (home appliances, sensors, actuators etc.) connected to each other via the internet, which produce a massive amount of data \cite{aboubakar2021review}. Consequently, data owners might be interested in sharing some of their data with other users, e.g. researchers, to analyse it and study its behaviour \cite{kamran2013formal}. As the need to share data increases, the importance of cloud services increases as well. 

\begin{figure}[H]
        \begin{center}
            \includegraphics[width = 5in, height = 2.29in,]{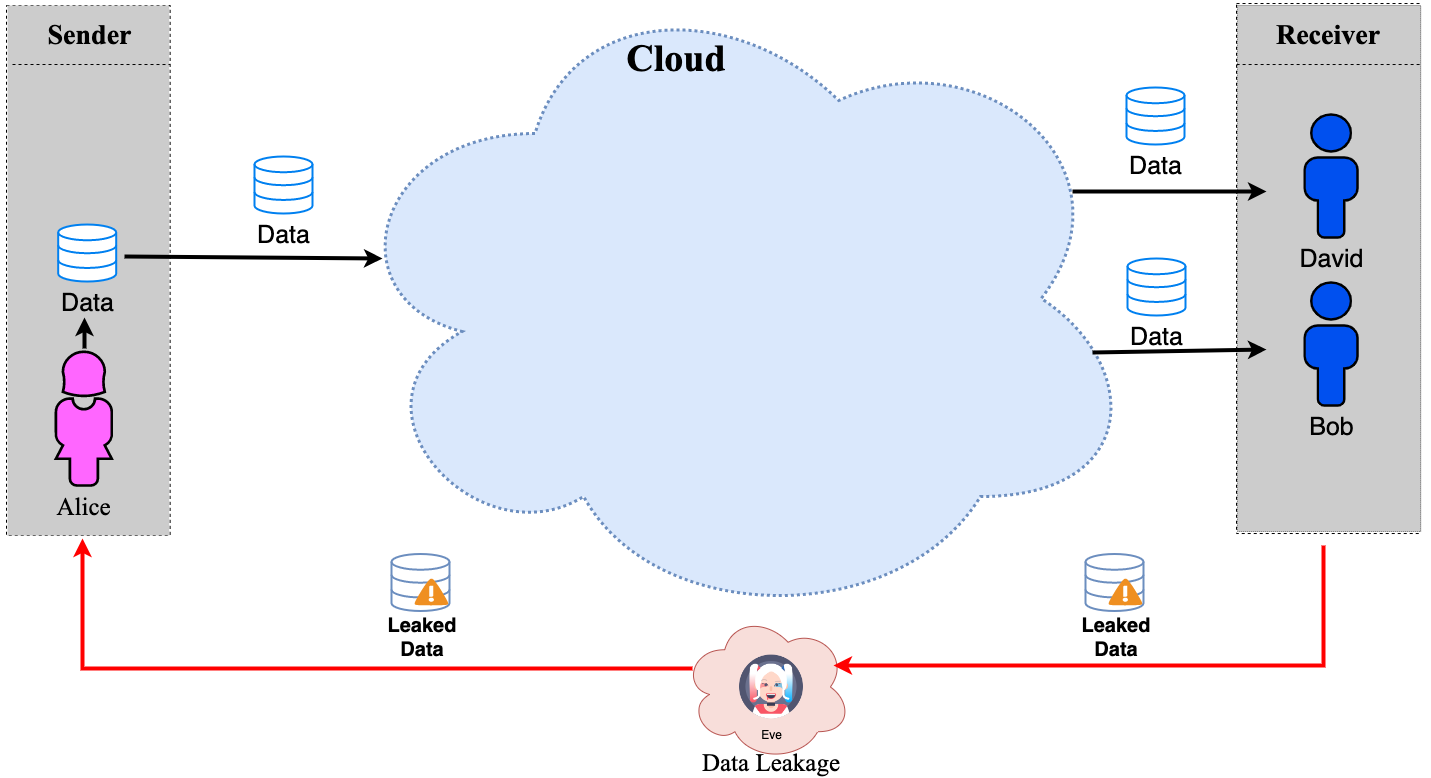}
            \caption{Data sharing scenario and stakeholders}
            \label{initialGraph}
        \end{center}
    \end{figure}

    The cloud is a set of hardware and resources centralised to provide efficient computational performance and storage via the internet \cite{alzakholi2020comparison}. Therefore, the cloud's role becomes more vital in collecting data from owners and distributing it to its recipients \cite{hussain2020lightweight}. This study considers the following stakeholders: Alice is the data owner, Bob and David are valid data recipients (researchers), the cloud is a third party (neutral judge), and Eve is an unauthorised user, as illustrated in Figure \ref{initialGraph}.
    
    Although sharing data with Bob could be beneficial for Alice because he could help her to better understand her data, sharing data may breach data privacy. In fact, data privacy can be breached in two main ways, either if Eve accesses the data directly or if Bob uses the data illegally, e.g. redistributes it to Eve without permission \cite{cox2007digital}. Eve can be prevented from directly accessing the data using techniques such as encryption, allowing access only to valid data users like Bob because he has been provided the decryption key. Whereas preventing Bob from behaving illegally with data he receives as a valid user is a more arduous problem. This is where digital watermarking becomes extremely useful \cite{cox2000watermarking}.
    
    Digital watermarking (DW) is an efficient technique to protect data ownership, data authentication, and more \cite{tao2014robust}. DW works by embedding a unique value (the watermark) into the data to serve the purpose that it embedded for \cite{Panah2019}. It is worth mentioning that the DW does not prevent illegal behaviour in relation to the data; however, it does give authorities the ability to track or identify such behaviour \cite{su1998digital}. For example, if Bob has distributed watermarked data to Eve, then the authorities will be able to use the watermark to identify Bob as the individual who supplied the data. In traditional watermarking systems, Alice, the data owner, would generate the watermark, embed it into a copy of the data, and then distribute the watermarked copy to Bob, which may not be fair to Bob \cite{kuribayashi2005fingerprinting}. 
    
    As Alice is the user who can generate, watermark and then distribute data, she may redistribute the same watermarked copy to different recipients either intentionally or unintentionally \cite{kuribayashi2005fingerprinting}. For example, if Alice watermarks a copy of data using Bob's watermark then distributes that copy to both Bob and David, and later that copy is found where it is not supposed to be, e.g. with Eve, Bob will be blamed for the breach because his watermark is on the data. However, it could have been David who actually redistributed that copy to Eve. In this case, although Bob is totally innocent, he will be one blamed, which is unfair to Bob \cite{frattolillo2006web}. To overcome this problem, various researchers have suggested an interactive watermarking approach \cite{kuribayashi2005fingerprinting}, \cite{pfitzmann1997anonymous}.  
    
    In the interactive watermarking system, the data is supposed to be watermarked by both Alice and Bob to ensure Bob will have his own unique watermark, which will be different to David's unique watermark \cite{pfitzmann1999coin}. Although the interactive watermarking approaches solve the problem of watermarking data by one user, they may not be practical enough in many cases.
    
    In fact, using the interactive watermarking approach introduces a couple of new issues \cite{frattolillo2006web}. For example, it requires a double watermark insertion by both Alice and Bob, which may affect the data quality because the second watermark may discredit the first watermark. Furthermore, it requires both Alice and Bob to have a digital certificate to verify their identity, whereas they may not own a digital certificate. Therefore, other researchers suggest watermarking the data using a third party \cite{zheng2011implementation}, \cite{zheng2012walsh}. 
    
\begin{figure}[H]
        \begin{center}
            \includegraphics[width = 5in, height = 3in,]{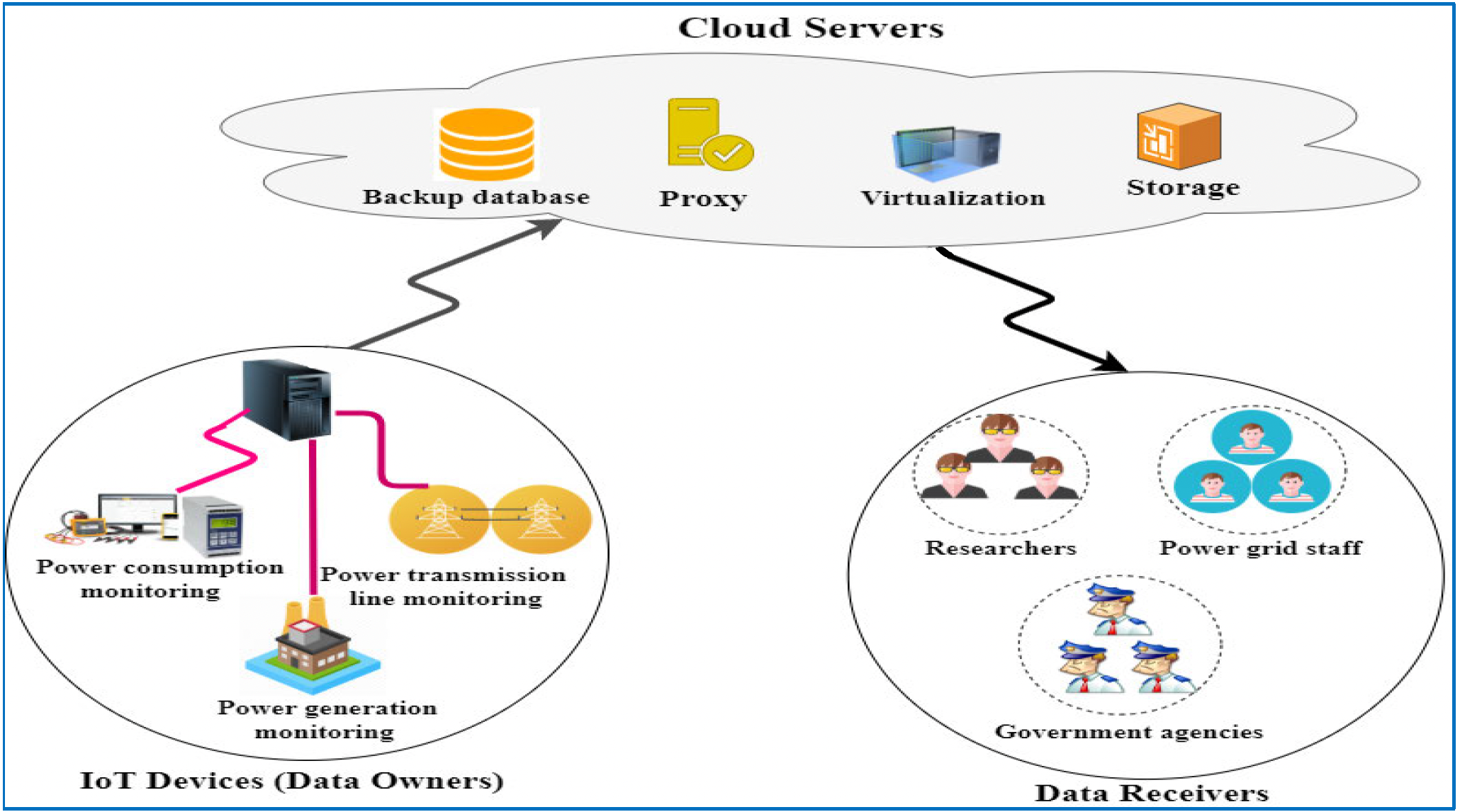}
            \caption{Modern systems infrastructure}
            \label{modernsysinfra}
        \end{center}
    \end{figure}

    Indeed, using a third party, e.g. the cloud, that is aligned to neither Alice nor Bob, to insert and verify the watermark will solve the majority of the previously raised problems \cite{frattolillo2006web}. Fortunately, in many current scenarios, the cloud already plays an essential role in linking users via the internet \cite{aboubakar2021review}. To clarify that, in many modern internet infrastructures, such as the presented in Figure \ref{modernsysinfra}, the data owners send their data to the cloud, and then the cloud redistributes it to the recipients \cite{hussain2020lightweight}. In addition, the cloud has strong computational power \cite{alzakholi2020comparison} which enables it to watermark data more efficiently compared to normal users, e.g. Alice, Bob, and David. Thus, the cloud seems a good potential candidate to undertake the responsibility of watermarking the data.
    
    It would also be beneficial if the cloud, as an impartial party, was delegated the responsibility to generate the watermark and then watermark the data with the generated watermark. In that case, if a copy of data is found in Eve's possession, the cloud will be the neutral judge to verify the data ownership and resolve the ownership conflict. However, the cloud should be able to embed the watermark without breaching data privacy, and also maintain the usefulness of the watermarked data.
    
    To enable the cloud to process the data while preserving its privacy, many researchers suggest using homomorphic encryption. Homomorphic encryption (HE) is a type of cryptography that allows arithmetic operations to be performed on encrypted data \cite{cheon2017homomorphic}. Indeed, many researchers already encrypt data using HE to allow the cloud to watermark it without breaching its privacy, but their implementations have some limitations.
    
    Two main limitations exist in the research that enables watermarking the encrypted data. The first limitation is that there is a considerable amount of noise introduced to the watermarked data due to the computational operations performed on the encrypted data \cite{zheng2011implementation}, \cite{zheng2012walsh}. This amount of noise may interfere with the usefulness of the watermarked data even for digital image data. In general, images are less sensitive to noise compared to other types of data, such as numerical data \cite{kamran2013formal}, which this research deals with. The second limitation is that many of those research studies use reversible watermarking algorithms \cite{li2020homomorphic}, \cite{liang2021robust}, \cite{li2019reversible}, \cite{xiang2018database}, \cite{zhang2021hope} in an attempt to preserve the data usefulness. The reversible watermarking algorithms are definitely suitable for some scenarios where Bob and David can remove the watermark and use the data, such as data authentication applications \cite{zhang2021hope}; however, the reversible algorithms may not be suitable for scenarios where Bob and David need be able to use the data while the watermark is embedded, such as copyright protection and source tracking applications \cite{kamran2013formal}.
    
    It is quite challenging to maintain the usefulness of numerical data after embedding a watermark \cite{kamran2013formal}. This becomes even more challenging if the data is going to be watermarked in an encrypted domain because performing arithmetic operations on encrypted data introduces extra noise \cite{zheng2011implementation}, \cite{zheng2012walsh}, \cite{lu2021pegasus}. Thus, one of the goals of this research is to determine the possibility of watermarking numerical data in an encrypted domain while maintaining the correctness of its statistical results using already existing tools and algorithms. 
    
    This research aims to overcome those issues by building a watermarking framework for ownership protection applications that enables the cloud to watermark encrypted univariate data using irreversible watermarking algorithms while controlling the usability of the watermarked data. The numerical data usability will be measured in terms of the number of statistical calculations that can be performed on the data that return the correct results \cite{kamran2013formal}.
    
    The framework will authorise the cloud to control the usability of the watermarked data by determining the correctness of the statistical calculations can be retrieved of the watermarked data. In other words, the framework will contain different irreversible watermarking algorithms, and each watermarking algorithm can preserve the accuracy of the results of specific statistical calculations, e.g. (\textit{[sum and mean] or [sum, mean, standard deviation and variance]}). Based on the selection of operations specified by Alice, the cloud will watermark the data using the suitable watermarking algorithm and then distribute the watermarked data to the intended recipient Bob. However, a watermark based on one user alone (e.g. the cloud) might not always be in the best interests of the recipient (e.g. Bob).

    According to Frattolillo and D'Onofrio \cite{frattolillo2006web}, if the watermark is embedded by one user (e.g. Alice or the cloud), then other users (e.g. Bob and David) should have the right to be confident that their copies are watermarked using watermarks related to them. To illustrate that, since the watermark is supposed to verify the owner's identity, the watermark must be unique and related to the recipient. That means every user (e.g. Bob and David) will know that their copies of data are watermarked with unique values known to them. Hence, if a copy of the data is leaked, the cloud can identify the original recipient via that unique value. However, although this step will provide compelling proof original recipient, it conflicts with a security aspect of the watermark \cite{cox2007digital}.  
    
    According to Cox et al. \cite{cox2007digital}, the security of the watermark can be evaluated from different aspects, one of which is unauthorised detection and verification of the watermark. Unauthorised verification of the watermark refers to the ability of unauthorised users to detect and verify the embedded watermark. Cox et al. \cite{cox2007digital} argue that any user who is able to verify the embedded watermark is probably able to remove it, which breaches the watermark security. 
    
   
   Therefore, Bob should know that the watermark has been embedded into his copy of the data to ensure it is bound to him \cite{frattolillo2006web}; however, he should not be able to verify the watermark by himself. For that reason, the security of the chosen watermarking algorithm should be reinforced so that the cloud becomes the only user that can verify the embedded watermark. In other words, the cloud will become the neutral judge in resolving data ownership conflicts, assuming that the cloud does not collude with any user. To illustrate, in case a copy of the data has been leaked, the cloud should be the only user that can verify the data ownership via the watermark even if other users (e.g. Bob and David) know the specific value that has been embedded in their data as a watermark, and this is the main contribution of this research. 
    
    This research contributes to creating a balance between Bob's right to know the watermarking value that has been embedded into his data, and to protect the security of the embedded watermark by preventing him from verifying it. In other words, it contributes by robustifying the security of verifying the watermark without affecting the data usability or the embedded watermark using a data obfuscation technique. Data obfuscation (DO) is a set of techniques used to decrease data sensitivity while maintaining its usefulness in various ways \cite{Bakken2004}. The chosen method makes the process of verifying the data ownership extremely difficult for Bob, even if he knows the actual value of the watermark. For instance, the time complexity for Bob to find the correct combination using a brute force attack is $\mathcal{O}(\frac{n !}{r !(n-r) !})$ based on the chosen watermarking algorithms. 
    
    In summary, the scenario \ref{initialGraph} on which this study is based revolves around Alice wanting to share her data with Bob, but in case the data is leaked to someone else (e.g. Eve), Alice will want to identify the user responsible for redistributing the data. Firstly, Alice encrypts the data homomorphically using Bob's public key. Secondly, she sends the encrypted data to the cloud, in addition to specifying the statistical calculations that should be maintained in the encrypted data. Thirdly, once the cloud receives the encrypted data, the cloud uses a secret watermark which is known only to the cloud and to Bob to watermark the encrypted data. After that, the cloud obfuscates the encrypted and watermarked data. Then, the cloud sends the obfuscated, watermarked, encrypted data to Bob. Finally, since the data was initially encrypted using Bob's public key, he can decrypt it using his private key. If Bob redistributes his copy to Eve, then the cloud will be able to determine his identity as the person who leaked the data via the watermark embedded in the data. Further details will be provided in the methodology chapter. 

\section{Motivation}
	\paragraph{}
	Digital watermarking is one of the most efficient methods used to restrict ownership violations and illegal data distribution. However, such applications require the watermark to be permanently embedded into the data. This research intends to advance the knowledge by creating a framework for ownership protection applications that allows a third party (the cloud) to be able to watermark numerical data using irreversible watermarking algorithms without revealing the actual content of data. The advantage of the proposed framework is the ability to give the cloud control over the usability of the watermarked data by guaranteeing the correctness of the results of particular statistical calculations performed on the data. Also, this research contributes to creating a balance between affording the recipient (Bob) the right to know the watermark value that is embedded in his data, and increasing the robustness of the security of the chosen watermarking algorithms whereby only the cloud can verify the watermark without affecting the watermark or the data usability. In other words, neither the sender (Alice) nor the recipient (Bob) gets involved in the watermarking or watermark verification processes, which means the cloud will be the only user that can verify the watermark embedded in the data even if the watermark is known. Specifically, the cloud will be the neutral judge in resolving data ownership conflicts.

\section{Research Questions}
	\paragraph{}
	 From the literature, it is evident there is a lack of research in which a third party (the cloud) watermarks numerical data using irreversible watermarking algorithms while preserving the result's accuracy for the desired statistical operations in encrypted domain. Therefore, this research aims to develop a framework that enables the cloud to watermark encrypted data using irreversible watermarking algorithms while maintaining the accuracy of the results of specific operations on the data. Furthermore, the framework will strengthen the security of the chosen watermarking algorithms by robustifying the process of verifying the ownership of the watermarked data, which is the main contribution of this research.
	 
    In order to construct such a framework, this research must address the following questions:
    
    \begin{itemize}
         \item \textit{Under what conditions it is possible to watermark numerical data using irreversible watermarking algorithms while maintaining the accuracy of the results of selected statistical operations in an encrypted domain? How accurate are the results of operations on data after watermarking it in an encrypted domain compared to those performed on data watermarked in an unencrypted domain? How efficient is it to watermark the data in an encrypted domain in terms of performance and memory consumption? }\\
         This question will be addressed using existing algorithms, platforms, and techniques.
         
        \item \textit{How can the security of the chosen algorithms be efficiently reinforced in terms of verifying the embedded watermark, making it extremely difficult for unauthorised users to verify it even if they know the watermark? What is the time complexity for unauthorised users to verify the watermark without the secret key?}\\
        The proposed method must affect neither the watermark nor the data usability for the selected statistics.
        
    \end{itemize}
    
    \section{Contribution}
        \paragraph{}
        Many researchers suggest watermarking encrypted data using reversible watermarking algorithms to maintain the data's usefulness. However, this research introduces the first implementation of a watermarking framework that can be used to watermark homomorphically encrypted data using an irreversible watermarking algorithm while maintaining the accuracy of the results of desired statistical operations performed on the watermarked data for ownership protection applications. Furthermore, the proposed framework nominates a third party (the cloud) to watermark the encrypted data and to be the neutral judge that can verify the ownership of the data in case it gets leaked. Therefore, the research contributes by strengthening watermark security via reinforcing the watermark verification process without affecting data usability. This contribution will create a balance between the recipient's right to know the watermark that has been embedded into their data, and the security of the watermark whereby only authorised users (the cloud) can verify the data ownership.
    
    \section{Organisation}
        \paragraph{}
        The remainder of the thesis is organised as follows. The second chapter reviews the literature on watermarking encrypted data, algorithms designed to watermark numerical data, encryption schemes and tools that can enable watermarking encrypted data, and obfuscation techniques to reinforce watermark verification security. The third chapter introduces the adopted methodology by describing the algorithms, tools, and techniques used to construct the framework, and also describes the proposed framework. The fourth chapter highlights the data used in this research, the experimental environment, and the settings used to successfully conduct the experiment. The fifth chapter demonstrates, analyses, and discusses the results that have been concluded from the research. The last chapter presents the conclusion, limitations, recommendations, and potential future work. 


\chapter{Literature Review}
\paragraph{}
    This thesis intends to construct a framework that will enable a third party (the cloud) to watermark numerical data using irreversible watermarking algorithms while preserving data accuracy and data privacy. The framework will also give the cloud the ability to control the usability of the watermarked data in terms of the statistical calculations which can be retrieved from the watermarked data. Finally, it will authorise the cloud to be the only user that is able to verify the ownership of the watermarked data. The aim of this chapter is to review relevant research studies, algorithms, platforms, and techniques to determine where the knowledge falls short, and fill the gaps.
    
    This chapter explores four main topics, each in a separate section. The first section investigates previous work related to \textit{watermarking encrypted data} where the data can be numerical, 2D or 3D images, aiming to identify the knowledge gap that this research intends to address. The second section reviews watermarking algorithms that are designed particularly for \textit{watermark numerical data}, and to review their features. The third section scrutinises different state-of-the-art \textit{encryption schemes} that could be adopted to achieve the research goals. The final section surveys \textit{data obfuscation techniques} that can be exploited to reinforce the security of the chosen watermarking algorithms without affecting the data usability, which is the main contribution of this research.
    
\section{Watermarking Encrypted Data}
    \paragraph{}
    Watermarking encrypted data has been a challenging problem for more than a decade. Many recent studies have focused on the possibility of allowing a third party (the cloud) to watermark various kinds of data such as 2D images, 3D images, and numerical data, without breaching data privacy. That means the watermarking process should be done in an encrypted domain. In fact, the majority of those studies assume that the data are encrypted using homomorphic encryption, so that it becomes possible to watermark the encrypted data. This section explores and reviews those studies and highlights their features and limitations. 
        
    Zheng and Huang \cite{zheng2011implementation} proposed a framework to use to watermark 2D images using the discrete wavelet transform tool (DWT) and multi-resolution analysis (MRA) in an encrypted domain. The researchers used the Paillier cryptosystem \cite{paillier1999public} to watermark the homomorphically encrypted data. Although they developed an improved multiplicative inverse method to reduce the data expansion error (caused by filter coefficient quantisation), their method still introduced an amount of error that may not be acceptable for some applications such as medical imaging \cite{zheng2012walsh}. Even though images are less vulnerable to being distorted by this introduced noise than other types of data, researchers have attempted various approaches to reduce the noise introduced by the watermarking process. 
        
    In one such approach to avoid the noise introduced in their previous study \cite{zheng2011implementation}, the same researchers suggested using the Walsh-Hadamard transform (WHT) algorithm to avoid the quantisation error \cite{zheng2012walsh}. The advantage of using the WHT algorithm is that its matrix contains only integer values. Another advantage is that it does not require multiplicative operations on the encrypted data, which makes it better suited for an additive homomorphic cryptosystem such as the Paillier cryptosystem \cite{paillier1999public}. However, Li et al. \cite{li2019reversible} suggested using a reversible watermarking method to watermark an image via grouping every couple of encrypted pixels together and then take the difference of them. After that, the watermark gets embedded by shifting the histogram of those differences. In their research \cite{li2019reversible}, they were able to embed the watermark into an encrypted image without the need to preprocessing the image before sending it. On the receiver's end, the user need to remove the watermark in order to restore the original image. In another attempt, Li et al. \cite{li2020homomorphic} introduced a reversible watermarking algorithm to insert a watermark into 3D models such as medical organs using the Paillier cryptosystem \cite{paillier1999public}. The researchers designed the algorithm to be more robust compared to the one used by Zheng and Huang \cite{zheng2012walsh}. They also tried to data usability by making the algorithm reversible in an attempt to preserve data quality. In their algorithm, the watermark is embedded by shifting the histogram. Later, Liang et al. \cite{liang2021robust} suggested a similar method to be applied to 2D images.
        
    In contrast, a few studies have been conducted on watermarking numerical data in an encrypted domain \cite{xiang2018database}, \cite{zhang2021hope}. In fact, studies have revealed that numerical data is much more sensitive to noise compared to other data types such as images \cite{kamran2013formal}. Therefore, many researchers have suggested reversible watermarking algorithms, so the watermark can be removed before using the data. For instance, Xiang et al. \cite{xiang2018database} introduced an algorithm to watermark encrypted databases for authentication purposes which used the watermark to detect whether or not the data had been tampered with. In their research, they used the order-preserving encryption (OPE) \cite{boldyreva2009order} scheme to enable them to perform comparison operations on encrypted data. A lossless watermarking method to watermark encrypted databases was proposed by Zhang et al. \cite{zhang2021hope}. In their scheme, the receiver must remove the embedded watermark before they use the data.  
        
    It is clear that, in an attempt to maintain data usability, many researchers have suggested using reversible watermarking algorithms on encrypted data to insert a watermark which should then be removed by the receiver before they use the data \cite{li2020homomorphic}, \cite{liang2021robust}, \cite{li2019reversible}, \cite{xiang2018database}, \cite{zhang2021hope}. This approach may not be suitable for applications where ownership verification is required. In fact, protecting the data ownership requires the watermark to be permanently embedded into the data, so that the ownership can be verified, especially if the data is leaked. At the same time, the usability of the watermarked data must be maintained, for example, for numerical data, the accuracy of the results of selected statistical operations needs to be preserved. 
        
    It can be concluded that there is a gap in knowledge regarding watermarking numerical data for ownership protection applications in an encrypted domain while maintaining the data usability. This research aims to fill that gap by creating a framework to use to watermark univariate encrypted data using different irreversible watermarking algorithms which preserves various statistical properties, using existing algorithms and tools. 
        
\section{Watermarking Numerical Data}
    \paragraph{}
    The problem of watermarking numerical datasets has attracted many researchers who have developed a number of watermarking algorithms designed specifically to watermark numerical data \cite{sion2002watermarking}, \cite{agrawal2003watermarking}, \cite{sebe2005noise}, \cite{sebe2006watermarking}. These algorithms have different features, such as the number of attacks that the watermark can survive or the usefulness of the data after the watermark is embedded. This section focuses on those studies and highlights their features.
        
    Sion et al. \cite{sion2002watermarking} introduced a non-blind watermarking algorithm specifically for watermarking numerical datasets. Then, Agrawal et al. \cite{agrawal2003watermarking} suggested a blind watermarking technique which is also designed for numeric databases. Both approaches \cite{sion2002watermarking} and \cite{agrawal2003watermarking} are robust against various attacks, including subset selection, subset resorting, and bit-flipping attacks. But, although the techniques in \cite{sion2002watermarking} and \cite{agrawal2003watermarking} are specifically designed to watermark numerical datasets, they do not guarantee the accuracy of the results of any statistical calculations subsequently performed on the watermarked data \cite{sebe2005noise}. 
        
    On the other hand, Sebe et al. \cite{sebe2005noise} developed a blind watermarking algorithm which is robust against additive noise attacks for univariate datasets. Their algorithm excels in comparison to those in \cite{sion2002watermarking} and \cite{agrawal2003watermarking} due to its ability to maintain the correctness of the results for certain statistical calculations such as mean and variance. It maintains the correctness of the mean and variance by generating particular parameters for each dataset separately via a nonlinear system of equations, which means the algorithm is nonlinear. As these parameters are valid only to watermark that particular dataset, the algorithm \cite{sebe2005noise} is considered to be a content dependent watermarking (CDW) algorithm. The same algorithm was later extended \cite{sebe2006watermarking} by Sebe et al. to handle multivariate datasets. In the extended version \cite{sebe2006watermarking}, the covariance is also maintained for the watermarked dataset. 
        
    In conclusion, this research aims to build a framework to use to watermark univariate datasets in an encrypted domain while preserving the correctness of the results for selected statistical operations. Those operations include [(mean) and (mean, variance)] for univariate datasets. The literature shows that the algorithms by \cite{sion2002watermarking} and \cite{agrawal2003watermarking} do not guarantee the correctness of the results for the watermarked data; consequently, they are not suitable to be adopted in this research. In contrast, as the Sebe 2005 algorithm \cite{sebe2005noise} can maintain the mean and variance for a univariate dataset, it is the most relevant algorithm to be adopted for use in this study. Whereas the Sebe 2006 algorithm \cite{sebe2006watermarking} is designed for multivariate datasets; therefore, it will be used in future work. The literature also shows a lack of algorithms designed to watermark numerical datasets to maintain only the mean. Therefore, this research intends to develop a simple watermarking algorithm similar to the one developed by Ibaida et al. \cite{ibaida2011low}. The simple algorithm is intended for illustrative purposes only and will not have security features. It will be capable of maintaining only the mean for numerical datasets, as will be demonstrated in the methodology chapter.
        
    The following table compares the three algorithms in terms of their features.
        
        \begin{table}[H]
    \footnotesize
\begin{tblr}{hlines, vlines,
             colspec= { X[1.4,l,m] *{7}{X[c,m]} X[c, m]},
             colsep = 3pt,
             row{1} = {font=\bfseries}
             }
 \SetCell[r=2]{m}  {Algorithm\\ name}
    &   \SetCell[c=3]{c}  {Statistical properties}
        &   &   &   \SetCell[c=4]{c}  {Algorithm's features}                                                                               
                    &   &   &   &   \SetCell[r=2]{m}  Notes     \\
    &   Average
        &   Variance
            &   Covariance 
                &   {Non-Blind}
                    &   Noise robustness
                        &   Linear algorithm
                            & CDW 
                                &                               \\
{Algorithm 1:\\ Simple algorithm}   
    & T & F & F & T & F & T & F & Illustrative purposes only    \\
Algorithm 2: Sebe 2005 \cite{sebe2005noise}  
    & T & T & F & F & T & F & T & None                          \\
Algorithm 3: Sebe 2006 \cite{sebe2006watermarking}  
    & T & T & T & F & T & F & T & Future work                   \\
\end{tblr}
    \end{table}

\section{Homomorphic Encryption Schemes}
	\paragraph{}
    As shown in Section 2.1, many researchers have used homomorphic encryption to encrypt data, so they can embed the watermark into the data while it is encrypted. Indeed, homomorphic encryption (HE) comes in handy in scenarios when an untrusted third party such as the cloud is required to process data without revealing it. In this research, the intention is for the cloud to be able to watermark encrypted numerical data using different watermarking algorithms, and one of these is the nonlinear algorithm \cite{sebe2005noise} introduced in the \textit{watermark numerical data} section. A nonlinear algorithm may require addition, multiplication, square root, or division operations on ciphertexts which may also result in decimal values. In other words, the adopted encryption platform must be able to deal with such circumstances. Therefore, this section explores the state-of-the-art libraries and platforms that can be used to permit the cloud to watermark the encrypted data using a nonlinear watermarking algorithm such as Sebe's algorithm \cite{sebe2005noise}. 
	
	Homomorphic schemes such as the Cheon, Kim, Kim and Song (CKKS) scheme \cite{cheon2017homomorphic} are considered to be word-wise HE schemes. Word-wise schemes support the efficient execution of polynomial functions, e.g. addition and multiplication \cite{smart2014fully}, but are inefficient when executing non-polynomial functions such as division and square root on the ciphertext. Whereas bit-wise HE schemes such as the fastest homomorphic encryption in the west (FHEW) \cite{ducas2015fhew} are much more efficient in executing non-polynomial functions but less efficient in executing polynomial functions \cite{cheon2019towards}. Since the Sebe Algorithm 1 \cite{sebe2005noise} requires polynomial and non-polynomial functions be performed on encrypted data, it is quite important to find an efficient platform that can execute both types of functions.
	
    Fortunately, there are a few platforms that allow polynomial and non-polynomial functions to be performed on ciphertext. The best candidate platforms are Chimera \cite{boura2020chimera} and Pegasus \cite{lu2021pegasus}. They both have similar functionalities in terms of operations that can be performed on the ciphertext, yet there are some differences in terms of efficiency. On the one hand, Chimera is designed to implement polynomial functions over both CKKS\cite{cheon2017homomorphic}/BFV\cite{fan2012somewhat} schemes, while implementing the non-polynomial functions on the fast fully homomorphic encryption scheme over the torus (TFHE) \cite{chillotti2020tfhe} using Look-Up Tables (LUT). On the other hand, Pegasus \cite{lu2021pegasus} uses only the CKKS \cite{cheon2017homomorphic} scheme which is able to deal with real numbers to implement polynomial functions, and uses FHEW \cite{ducas2015fhew} to implement non-polynomial functions using LUTs as well. According to Boura et al. \cite{boura2020chimera}, the key size that used in their conversion algorithm in Chimera is more than 10 GB. However, the key size for the conversion algorithm used in Pegasus \cite{lu2021pegasus} is roughly only one GB. In addition, Pegasus is considered to be much more efficient in terms of performance \cite{lu2021pegasus}.
    
    In conclusion, it is clear that the Pegasus platform is much more efficient in performing polynomial and non-polynomial functions on ciphertext compared to the Chimera platform. Therefore, for this research, Pegasus was chosen as the platform to use to encrypt data. 
	
\section{Data Obfuscation Techniques}
	\paragraph{}
    Data obfuscation (DO) is a privacy-preserving technique widely used to reduce data sensitivity while maintaining its usability \cite{Bakken2004}. It is heavily used to preserve the privacy of individual data points in scenarios such as data mining. There are different forms and classifications of DO based on its purpose. This research requires a technique that is able to strengthen the watermark security by making the process of verifying the data ownership extremely difficult without affecting the data usability. Thus, this section investigates various kinds of DO techniques and highlights their usage in an attempt to identify a suitable technique.
    	
    Tonyali et al. \cite{Tonyali2016} used DO to preserve the privacy of consumer usage data. In their research, the DO was mainly used to anonymise individual records while allowing the receiver the ability to perform statistical calculations on the data. A survey of different DO techniques was conducted by Seidl et al. \cite{Seidl2015} where they investigated three DO techniques to find a balance between releasing spatial data and protecting personal identities stored in the dataset. They used grid masking, random perturbation, and weighted random perturbation techniques and evaluated the effectiveness of each technique in obfuscating the data. Furthermore, they introduced a data masking technique that is more efficient than the other three. A reversible DO technique was also used by Panah et al. \cite{Panah2019} to blur critical data to different degrees, based on the privilege of the receiver. Another reversible DO technique was used by Hessler et al. \cite{Hessler2012} to obfuscate and conceal streaming payload data during its transmission via a network to protect it from eavesdroppers. 
    	
    A different kind of DO technique, the swapping technique, was introduced by Gomatam and Karr \cite{gomatam2003distortion}. Their study compared and contrasted the data swapping technique in terms of categorical approaches. In general, the data swapping technique can be defined as the process of switching the position of the values in the same column. In other words, the data swapping technique does not change the value of each data point; however, it changes its sequence \cite{Bakken2004}. The purpose of the data swapping technique is to break the sequence of the data and de-link the values in each record in an attempt to reduce data sensitivity. This technique also ensures the correctness of the results of mathematical operations performed on the data such as mean, variance, median, mode, max, and others.
    	
    In conclusion, it is clear that various DO techniques are widely used in different forms and for various purposes. As this research requires an efficient technique to reinforce the security of the chosen watermarking algorithms without affecting the data usability and the embedded watermark, the DO technique chosen was the swapping technique \cite{Bakken2004} \cite{gomatam2003distortion}, which shuffles the data elements in each column separately. The DO scheme will contribute by reinforcing the watermark verification process. This will be discussed in the methodology chapter.

\chapter{Methodology}
\paragraph{}
   This research aims to build a framework that will enable a third party (the cloud) to watermark numerical data while controlling the usefulness of the watermarked data without breaching data privacy. In addition, the framework will reinforce the security of the process of verifying the ownership of watermarked data using an efficient technique. This chapter will present the algorithms, platforms, and techniques that will be used to construct such a framework. 
   
   The chapter is divided into four sections. The first section introduces the topic of \textit{digital watermarking} and algorithms that have been adopted in the framework. The second section contains a brief overview of \textit{homomorphic encryption} and the platform used by the framework to secure the data. The third section presents the \textit{data obfuscation} scheme, its categories and properties, and how it could increase the robustness of the security in the watermark verification process. The final section demonstrates the \textit{proposed framework} and how all of these techniques and algorithms will work together.

   \section{Digital watermarking (DW)}
        \subsection{Introduction}
        \paragraph{}
        The rapid acceleration of technology and the pressing need to share data between various users over the internet had raised significant issues related to data privacy. One efficient technique used to protect data from unauthorised users is cryptography; however, it does not protect data from illegal use by authorised users. For instance, some users who have the right to access certain data may illegally manipulate the data or distribute the data to unauthorised users, and this is where power of digital watermarking can be demonstrated \cite{su1998digital}.
            
        The concept of digital watermarking (DW) started attracting researchers' attention in the mid 1990s \cite{cox2007digital}. Since then, it has been widely applied to a variety of digital content such as images, videos, audios, numerical data. DW can be defined as the process of embedding certain information (the watermark) into digital content (data) without interfering with the purpose that data is supposed to serve. In other words, the usability of the watermarked data should be maintained even after embedding the watermark. Note that DW does not prevent the illegal distribution or manipulation of data; however, it can enable authorities to identify instances of such behaviour, which can help to reduce them \cite{su1998digital}. 
            
        \begin{figure}[h]
            \centering
        	\includegraphics[width = 6in, height = 1.3in]{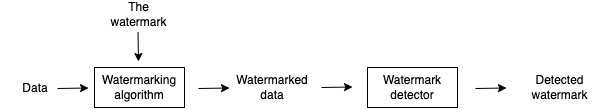}
        	\caption{High level watermarking system}
        	\label{HighLvlWTRSys}
        \end{figure}
            
        Figure \ref{HighLvlWTRSys} illustrates a high-level watermarking system that consists of two main stages \cite{cox2007digital}. The first stage is the watermark embedding process in which the watermarking algorithm takes two inputs, the data and the watermark, and embeds the watermark into the data. The second stage is the watermark detection process whereby the detection algorithm takes the watermarked data as an input and detects the embedded watermark \cite{cox2007digital}. To detect the watermark, the detection algorithm requires either the original data or a secret key, depending on whether the watermarking algorithm is a blind or non-blind algorithm \cite{tao2014robust}. This will be explained in the next section. 

            
        \subsection{Watermarking algorithm characteristics}
        \paragraph{}
        DW can be used to serve many purposes and applications, e.g. ownership protection, data authentication, tamper proofing and others. Hence, the characteristics of different watermarking algorithms may differ based on the specific application for which they were designed \cite{tao2014robust}. Practically, it is not possible for a single watermarking algorithm to demonstrate be superior in all characteristics; thus, there is usually a trade-off between certain characteristics \cite{cox2000watermarking}. This section highlights the most important characteristics of watermarking algorithms \cite{cox2007digital}, \cite{cox2000watermarking}, \cite{tao2014robust}, \cite{su1998digital}.
            
        \begin{itemize}
            
        \item \textbf{Robustness} means the watermark's ability to survive after unintentional or intentional processing of the data \cite{cox2007digital}, \cite{tao2014robust}. Unintentional processing may include alteration of the data due to a transmission process that compress the data. While intentional processing refers to the attacks launched by an adversary (e.g. noise attacks) on the watermarked data in an attempt to destroy the watermark. In both cases, the watermark is supposed to survive as long as the data is still usable. Although robustness is one of the most common and desirable characteristics of watermarking algorithms, there are some applications where robustness is not required or is even undesirable, such as authentication applications \cite{cox2007digital}. 
            
        \item \textbf{Imperceptibility} refers to the effect on or distortion of the original data due to the watermarking process \cite{cox2007digital}. Expressly, it is the possibility of perceiving the impact of the embedded watermark on the data based on the purpose the watermarked data is supposed to serve. Note that no watermark is imperceptible for the computer; however, it is possible to make the computer unsure whether the watermark is embedded. One of the main reasons for the importance of imperceptibility is that the watermark should not affect the purpose of using the data \cite{tao2014robust}. In the ideal case, the original and watermarked data should be identical \cite{su1998digital}.
            
        \item \textbf{Capacity} indicates the amount of information that can be embedded into the data as a watermark without compromising the robustness or imperceptibility of the watermark \cite{tao2014robust}. The watermark capacity might differ based on the type of data to be watermarked \cite{cox2007digital}. To illustrate, the capacity for watermarking a video may refer to the number of bits that can be embedded within a frame or into a second. Similarly, in audio watermarking, the capacity may indicate the number of bits that can be embedded into a second. Whereas for images, the capacity refers to the number of bits that can be embedded into a pixel. Generally, as long as the watermarking algorithm is able to embed the necessary information that serves its purpose, then its capacity is considered to be sufficient \cite{cox2007digital}. For example, in copyright scenarios, the watermarking algorithm should be able to embed enough information to identify and verify the identity of the data's owner.
            
        \item \textbf{Security} refers to how secure the watermark is against unauthorised users. The security of a watermarking algorithm can be evaluated from different aspects, one of which is the ability of unauthorised users to detect or verify the watermark \cite{cox2007digital}. In fact, it is essential for watermarking algorithms used in applications such as tracking illegal data distribution, copyright protection, and others, to be secure against such a vulnerability \cite{tao2014robust}. According to Cox et al. \cite{cox2007digital}, users who can detect or verify the watermark probably can remove it. Therefore, verifying the data's ownership via the watermark should be computationally extremely difficult for unauthorised users. On this occasion, this research contributes by reinforcing the security of the chosen watermarking algorithms via reinforcing the process of verifying the watermark. It uses an efficient method to ensure the process of verifying the data's ownership is exceedingly difficult for unauthorised users even if they own the watermark and the watermarked data. 
            
        \item \textbf{Fast embedding and retrieval} refers to how efficient the algorithm is in terms of embedding the watermark into the data and extracting/verifying the watermarked data. Although the efficiency of the watermark embedding process is quite critical especially for streaming data applications, Su et al. \cite{su1998digital} emphasise that watermarking algorithms should prioritise detection reliability over performance.
            
        \item \textbf{Blind or non-blind watermarking algorithms} indicates whether the algorithm requires a copy of the original data to detect or extract the watermark \cite{cox2007digital}. A blind algorithm does not require the original data to extract or verify the embedded watermark. This kind of algorithm is useful in many scenarios, especially when the original data is meant to be private such as in copyright protection applications \cite{su1998digital}. In contrast, a non-blind watermarking algorithm does require the original data to verify or extract the watermark. This kind of algorithm might be beneficial in scenarios where both the sender and receiver own the original data, for example, when both sender and recipient agree on a particular kind of data (e.g a specific image) and then embed secret information into that data and exchange the data to pass on the secret information \cite{cox2007digital}. In that case, both sender and receiver possess a copy of the original data. Another scenario is where the data owner is the user who is supposed to verify the watermark.
        \end{itemize}
            
        \subsection{Watermarking algorithms applications}
        \paragraph{}
            DW excels in comparison to other methods due to its flexibility in supporting various applications with different requirements. There are many applications that DW can efficiently support; however, due to the scope of this thesis, in this subsection the focus is limited to three applications.
            
            \begin{itemize}
             \item \textit{Data authentication} applications require the watermark to identify whether the data has been tampered with and, probably where the alteration occurred \cite{cox2000watermarking}. Any change made to the data will be reflected in the embedded watermark; consequently, that change will be detectable. In this scenario, a watermark is preferable to other techniques because it will be merged with the data to form an integral part of the data.
            
            \item \textit{Copyright protection} applications use the watermark to prove the identity of the copyright owner \cite{tao2014robust}. The embedded watermark should identify the user who owns the copyright of the data. In this scenario, all recipients receive the same copy of the data, and if another user claims ownership of the data, then the watermark should reveal the rightful owner.
            
            \item \textit{Source tracking (fingerprinting)} applications require the watermark to prove the identity of the recipient \cite{cox2000watermarking}, \cite{tao2014robust}. Each copy of the data contains a unique watermark that refers to a certain recipient. In this scenario, each recipient receives a copy of the data that contains a unique watermark that refers to their identity, and if a user leaks their copy of the data, then the watermark should be able to reveal the identity of the user who leaked the data. Source tracking applications require the watermarking algorithm to be robust against noise and forgery attacks. Also, the watermark must be embedded in data permanently.
            \end{itemize}
            
    \subsection{Watermarking algorithm attacks}
    \paragraph{}
    As with any privacy-preserving technique, an adversary can launch attacks against watermarked data. The severity of those attacks may differ based on the purpose and the application the watermark is supposed to serve. This section highlights the most common attacks \cite{cox2000watermarking}.
            
    \begin{itemize}
        \item \textbf{Active attacks} in these kind of attacks, the adversary attempts to illuminate the embedded watermark or make it undetectable (e.g. by adding extra noise). This kind of attack is one of the most critical for many applications, including source tracking applications.
                
        \item \textbf{Passive attacks:} in these kind of attacks, the adversary does not attempt to illuminate the watermark; instead, they attempt to verify the embedded watermark. According to Cox et al. \cite{cox2007digital}, whoever is able to verify the watermark is probably able to remove it. This thesis contributes by increasing the robustness of the chosen watermarking algorithms against this type of attack against unauthorised users, even if they own the watermark and the watermarked data.
                
        \item \textbf{Collusion attacks:} in these kind of attacks it is assumed that the adversary owns differently watermarked copies of the data and they attempt to reconstruct a new copy of the data which does not contain a watermark. This attack is considered a variation of the active attack.
                
        \item \textbf{Forgery attacks:} this attack represents an adversary's attempt to insert a fake watermark into already watermarked data. Consequently, the watermark verification algorithm will confuse between the real and fake watermarks. This attack is quite critical for many applications, especially copyright protection applications because more than one user may claim data ownership.
        
        \item \textbf{Estimation attacks:} in these kind of attacks, the adversary tries to estimate the watermark by analysing different copies of data which watermarked using the exact watermark. For this matter, the adversary collects many copies of data and they try to estimate it so that they can remove it.  
        \end{itemize}
            
    \subsection{Adopted algorithms}
    \paragraph{}
    One of the aims of this research is to build a framework that enables a third party to watermark encrypted data while maintaining the accuracy of the results of selected statistical operations, such as sum, mean, standard deviation, and variance. The framework will contain multiple watermarking algorithms, each intended to preserve specific statistical properties of data. This section illustrates two algorithms that were adopted along with some of their features.
            
    \subsubsection{Algorithm 1}          
    \paragraph{}
        The watermarking algorithm \textit{Algorithm 1} is supposed to preserve the sum and mean of the watermarked data. Unfortunately, the literature revealed a lack of watermarking algorithms designed for numerical data to preserve only the sum and mean. Therefore, in this study a simple watermarking algorithm similar to the algorithm used by Ibaida et al. \cite{ibaida2011low} was developed; however, this algorithm is for illustrative purposes only. 
                
        Algorithm 1 is a non-blind watermarking algorithm which means it requires the original data to verify the watermark. This algorithm, however, is not robust against noise attacks. In addition, it is designed to preserve only the sum and mean. Note that preserving the mean of the watermarked data also results in the sum being preserved.
                
        To preserve the sum and mean for the data $X$, the algorithm must maintain the same results for these statistical calculations run on the watermarked data $X^{\prime}$ as for the original data, as indicated by:
                
        \begin{equation}
            \label{eqn:preserveMean}
            \overline{X} = \overline{X^{\prime}}
        \end{equation}
                
        This algorithm consists of three stages: watermark generation, watermarking the data, and watermark verification. 
        
        \begin{itemize} 
        \item[] 
        \textit{\textbf{Watermark generation}}\\
            The watermark gets generated in the watermark generation stage. In this algorithm, the watermark is generated using a pseudorandom number generator. 
                
        The pseudorandom number generator (PRNG) produces a sequence of numbers based on a specific seed \cite{agrawal2003watermarking}. In general, the PRNG $G$ is deterministic property, meaning it will repeatedly generate the same sequence using the same seed. For this algorithm, the cloud chooses the seed (the watermark) $K$ for each user, which is a secret value known only to the cloud and the recipient.
                
        To satisfy the mean property highlighted in Equation \ref{eqn:preserveMean}, the algorithm generates watermarking values which sum up to zero.
                
        The pseudocode in Algorithm \ref{WatermarkGenerationAlgo1} describes the procedure followed to generate the watermark.\\
        
        \renewcommand{\thealgorithm}{1A}

        \begin{algorithm}
        \caption{Watermark generation stage}\label{WatermarkGenerationAlgo1}
                
        \begin{algorithmic}[0]
        \State \textbf{Input:} Input to the function is the watermark $K$ to be used as the seed by the PRNG, the secret key $SH_{key}$ for shuffling, and the length $n$ which is the required length of the watermark.
        \State \textbf{Output:} the watermarking sequence $S[n]$ is returned as an output.
                
        \begin{algorithmic}[1]
        \Function{WatermarkGeneration}{}
            \State Initialise PartialS[n/2]; \textit{// to store the positive half of the watermark sequence.}
            \State  Initialise S[n]; \textit{// to store the full watermark sequence.}
                
            \State PartialS[i] $\gets$ PRNG(K); \textit{// generate the positive half of the watermark.}
            \For{ int i $\rightarrow$ 0 to halfN} \textit{// generate the complement of each element. }
                 \State S[i] $\gets$ PartialS[i];
                \State S[i + halfN] $\gets$ PartialS[i] * (-1);
                
            \EndFor
            \State Obfuscate(S[n], $SH_{key}$) \textit{// shuffle the elements in $S[n]$ using the secret key $SH_{key}$.}
        \EndFunction
        \end{algorithmic}
        \end{algorithmic}
        \end{algorithm}

        \item[] 
        \textit{\textbf{Watermarking the data}}\\
        The watermarking stage is when the watermark gets embedded into the data. This stage requires two main elements: the dataset $X$ that is to be watermarked and the watermarking sequence $S[n]$ which was generated in the previous stage.
                 
        The pseudocode in Algorithm \ref{WatermarkingAlgo1} presents the procedure followed to watermark the dataset.\\
        
        \renewcommand{\thealgorithm}{1B} 
           
        \begin{algorithm}[H]
        \caption{Data watermarking stage}\label{WatermarkingAlgo1}
                    
        \begin{algorithmic}[0]
        \State \textbf{Input:} Input to the function is the original data $X[n]$, and the watermarking sequence $S[n]$. 
        \State \textbf{Output:} The watermarked data $X^\prime[n]$ is returned as an output.
        \begin{algorithmic}[1]
        \Function{Watermarking}{}
            \State Initialise $X^\prime[n]$
            \For{int i $\rightarrow$ 0 to n}
                \State  $X^\prime$[i]  $\gets$ X[i] + S[i]
            \EndFor
        \EndFunction
        \end{algorithmic}
        \end{algorithmic}
        \end{algorithm}
                    
        \item[] 
        \textit{\textbf{Watermark verification}}\\
        The watermark verification stage is where the watermark in the watermarked dataset $X^\prime[n]$ is detected and verified. This stage is the reverse of the watermarking stage but with different inputs.
                 
        As mentioned previously, \textit{Algorithm 1} is a non-blind watermarking algorithm, which means it requires the original data to extract the watermark. Therefore, this stage requires two inputs, which are the original dataset $X$ and the dataset which expected to be watermarked $\hat{X}$. 
                 
        The pseudocode in Algorithm \ref{VerifyingAlgo1} shows the process of verifying the watermark.\\

        \renewcommand{\thealgorithm}{1C}       
        \begin{algorithm}
        \caption{Watermark verification stage} \label{VerifyingAlgo1}
                    
        \begin{algorithmic}[0]
        \State \textbf{Input:} Input to function is the original data $X[n]$, and the watermarked data $\hat{X}[n]$.
        \State \textbf{Output:} The detected watermark $S[n]$ is returned as an output.
        \begin{algorithmic}[1]
        \Function{WatermarkDetection}{}
            \For{int i $\rightarrow$ 0 to n}
                \State  S[i] $\gets$ X[i] - $\hat{X}$[i]
            \EndFor
        \EndFunction
        \end{algorithmic}
        \end{algorithmic}
        \end{algorithm}
    \end{itemize} 
    
    \subsubsection{Algorithm 2}
    \paragraph{}
    The watermarking algorithm \textit{Algorithm 2} is supposed to preserve the sum, mean, variance, and standard deviation in the watermarked data. Fortunately, an existing watermarking algorithm was found that was designed specifically to preserve the sum, mean, variance, and standard deviation, developed by Sebe et al. \cite{sebe2005noise}. This algorithm \cite{sebe2005noise} has been adopted for use in this research.
                
    Sebe's algorithm is explicitly designed to watermark numerical data. Being a blind watermarking algorithm, it does not require the original data to verify the watermark. Also, it is robust against noise, forgery, and estimation attacks.
                
    To preserve the sum and mean for the original data $X$, the algorithm must maintain the same results for statistical calculations run on the watermarked data $X^{\prime}$, as illustrated by Equation \ref{eqn:preserveMean}.
                
    It is also worth mentioning that preserving the variance in the watermarked data leads to preserving the standard deviation. To preserve the variance and standard deviation in the original data $X$, the algorithm must maintain the same results of statistical calculations on the watermarked data $X^{\prime}$, as stated by Equation \ref{eqn:preserveVariance}.
                
        \begin{equation}
        \label{eqn:preserveVariance}
        S_{X}^{2} = S_{X^{\prime}}^{2}
        \end{equation}
                
    This algorithm consists of three stages: constant generation, data watermarking, and watermark verification.
    
     \begin{itemize} 
        \item[] 
        \textit{\textbf{Constant generation}}\\
        Sebe's algorithm is considered to be a content-dependent watermarking (CDW) algorithm which means it depends on the data to generate the watermark. This feature makes the algorithm more robust against some attacks, such as estimation attacks \cite{li2007constructing}. A CDW algorithm is able to generate different watermark values for each dataset. This kind of algorithm requires specific values to be generated from each dataset. Sebe's algorithm specifically requires the mean and variance from each dataset, as will be illustrated in this section. 
                
        Initial values that Sebe's algorithm requires to be generated are as follows:
        \begin{enumerate}
            \item Using the PRNG $G$ which is seeded by the watermark $K$ to generate the sequence $S=\{s_{1}, s_{2}, ..., s_{n}\} $, where $s_{i}$ = \{-1, 1\}, and $\probP(s_{i} = -1)$ = $\probP(s_{i} = 1) = 1/2 $.\\
            \item Using the PRNG $G$ which is seeded by a random value $R$, generate the sequence $T=\{t_{1}, t_{2}, ..., t_{n}\}$ where the sequence follows a Gaussian $N(0,1)$ distribution.\\
            \item Choose the value for the parameter $M$, which should be kept secret by the watermarker (the cloud). As the value of $M$ gets larger, the value of $\epsilon$ gets smaller. The parameter $M$ used to verify the existence of the embedded watermark.
            The value of $\epsilon$ determines the error in the results of the watermarked dataset. In other words, the results' accuracy increases as $\epsilon$ gets smaller. It can be calculated using formula \ref{eqn:calculateepsilon}.
                
            \begin{equation}
                \label{eqn:calculateepsilon}
                \epsilon = \frac{2 E\left[X^{2}\right]}{M^{2} n}
            \end{equation}
            \\ 
            \item Compute the following nonlinear system of equations \ref{eqn:firstEquation}, \ref{eqn:secondEquation}, and \ref{eqn:thirdEquation} to generate the three parameters a, b, and $\lambda$. These three parameters will be used in the watermarking stage. In this step, the algorithm requires the mean $\overline{X}$ and variance $S_{X}^{2}$ for each dataset to be watermarked, which is why Sebe's algorithm is classified as a CDW algorithm.
                
            \begin{equation}
                \label{eqn:firstEquation}
                \overline{X}=a \overline{X}+b+\lambda \overline{S|T|} 
            \end{equation}
                
            \begin{equation}
                \label{eqn:secondEquation}
                S_{X}^{2}=a^{2} S_{X}^{2}+\lambda^{2} S_{S|T|}^{2}
            \end{equation}

            \begin{equation}
                \label{eqn:thirdEquation}
                M=a \overline{X S}+b \overline{S}+\lambda \overline{|T|}
            \end{equation}
        \end{enumerate}
        
        In order to compute this system of equation, it will require variety of arithmetic operations such as addition, multiplication, square root and others. 
                
        Finally, most of the values that have been generated in this stage, such as $S,$ $T,$ $a,$ $b,$ and $\lambda$, will be passed to the next stage to watermark the dataset.  
        
        \item[] 
        \textit{\textbf{Data watermarking}}\\
        To watermark a dataset, Sebe's algorithm requires a linear equation to be implemented which watermarks each element in the data. Equation \ref{eqn:watermarking} requires some of the parameters that were generated in the previous stage such as $S,$ $T,$ $a,$ $b,$ and $\lambda$, as follows:
                
        \begin{equation}
            \label{eqn:watermarking}
            x_{i}^{\prime}=a x_{i}+b+s_{i}\left|t_{i}\right| \lambda
        \end{equation}
                
        Watermarking equation \ref{eqn:watermarking} ensures that each element in the data $X$ is watermarked in the data ${X^{\prime}}$. Also, it requires performing both addition and multiplication operations for each element in $X$.
                 
        \item[] 
        \textit{\textbf{Watermark verification}}\\
        The watermark verification stage is where the algorithm verifies the watermark embedded in the data. This stage requires two inputs: the data $\hat{X}$ which is assumed to be watermarked, and the watermark $K$. The verification procedure is as follows:
                
        \begin{enumerate}
            \item Using the PRNG $G$ which is seeded by $K$, generate the sequence $S=\{s_{1}, s_{2}, ..., s_{n}\} $, where $s_{i}$ = \{-1, 1\}, and $\probP(s_{i} = -1)$ = $\probP(s_{i} = 1) = 1/2 $.
                
            \item Calculate $\hat{M}$ as illustrated in Equation \ref{eqn:watermarkVerification}.
                
            \begin{equation}
          \label{eqn:watermarkVerification}
                \hat{M}=\frac{1}{n} \sum_{i=1}^{n} s_{i} \hat{x}_{i}
            \end{equation}
        \end{enumerate}
                
        From Equation \ref{eqn:watermarkVerification}, this stage returns whether the dataset $\hat{X}$ is watermarked using the watermark $K$ or not, based on the following conditions:
                
        \begin{center}
         $\left\{\begin{array}{l}\operatorname{If} \hat{M}>\frac{M}{2} \rightarrow \hat{X} \text { is watermarked by }  K.  \\ \operatorname{If} \hat{M} \leq \frac{M}{2} \rightarrow \hat{X} \text { is not watermarked by }  K. \end{array}\right.$
        \end{center}
        \end{itemize} 
 
        
        
               
    \section{Homomorphic encryption}
    \subsection{Introduction}
        \paragraph{}
        Homomorphic encryption (HE) is a type of cryptography that allows arithmetic operations to be performed on the encrypted data without the need to decrypt it first. For a cryptosystem to be described as having homomorphic properties, it must maintain the following:
        
        \begin{equation}
            \label{eqn:homomorphicEquation}
            E\left(x_{1}\right) \star E\left(x_{2}\right)=E\left(x_{1} \star x_{2}\right), \quad \forall \text{ } x_{1}, x_{2} \in X
        \end{equation}

        
        \noindent where $E(.)$ is the encryption function, and $\star$ is a certain arithmetic operation (e.g. addition), while $X$ represents the set of potential messages \cite{Acar2018}. Adding two encrypted numbers should be equal to adding two unencrypted numbers, except that adding two encrypted numbers will give an encrypted result, which requires a secret key to decrypt.
        
        The idea of using homomorphic encryption to secure data was first proposed by Rivest et al. \cite{rivest1978data} in 1978. Then in 2009, Gentry \cite{gentry2009fully} introduced the first fully homomorphic framework. Since then, many HE libraries and frameworks have been developed using different schemes \cite{halevi2014algorithms}, \cite{laine2016simple}.
        
         HE cryptosystems differ from other cryptosystems such as Advanced Encryption Standard (AES) cryptosystem \cite{daemen2001reijndael} in many aspects. One of these is that many HE cryptosystems are believed to be secure against both classical and quantum computers, especially those based on the Ring Learning-with-Errors problem. The Ring Learning-with-Errors (RLWE) problem is a variation of the Learning-with-Errors problem, which is a mathematical problem that is assumed to be difficult to be solved using classical and quantum computers under certain parameters \cite{lauter2021protecting}. Another aspect is that, while the HE scheme can be used as an asymmetric cryptosystem where it requires one key (the public key) to encrypt a message and another key (the secret key) to decrypt the message, it can also be used as a symmetric cryptosystem where only one key is required to both encrypt and decrypt messages \cite{Acar2018}. Furthermore, HE has advantages over other cryptosystems because it allows computations on the ciphertext, which is the main reason for using it in this research. 
         
\begin{figure}[H]
            \begin{center}
                \includegraphics[width = 3in, height = 3in,]{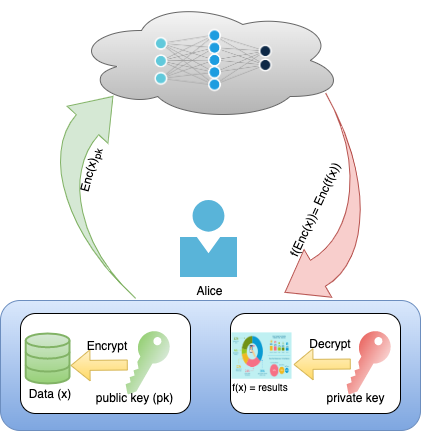}
                \caption{An example of homomorphic applications}
                \label{homoExample}
             \end{center}
        \end{figure}

         Actually, HE is widely used in applications that require a third party (the cloud) to perform specific computable functions on encrypted data without revealing the actual content of data \cite{cheon2017homomorphic}. One example of using HE is shown in Figure \ref{homoExample}, whereby the cloud owns a private machine learning (ML) model which it does not want to expose it to the public; on the other side, the data owner wants to exploit the benefits of the ML model without breaching data privacy \cite{pham2021private}.
         
        \subsection{Homomorphic encryption categories}
        \paragraph{}
        There are plenty of homomorphic algorithms and schemes with different characteristics that vary in terms of the calculations that can be performed and the number of times those calculations can be performed on the encrypted data. Homomorphic schemes can be categorised into partial homomorphic encryption, somewhat homomorphic encryption, or fully homomorphic encryption.
        
        \begin{itemize}
            \item \textbf{Partial homomorphic encryption} refers to cryptosystems that allow a specific operation to be performed (e.g. addition or multiplication) an unlimited number of times. For example, Paillier cryptosystem \cite{paillier1999public} has the additive homomorphism property, while Elgamal cryptosystem \cite{elgamal1985public} has the multiplicative homomorphism property.  
            
            \item \textbf{Somewhat homomorphic encryption} indicates cryptosystems that allow both addition and multiplication to be performed on encrypted data with some limitations. For instance, the Boneh-Goh-Nissim (BGN) cryptosystem \cite{boneh2005evaluating} allows an unlimited number of additions to be performed, but it allows only one multiplicative operation to be run on the ciphertext.

            \item \textbf{Fully homomorphic encryption} describes cryptosystems that support an unlimited number of operations (e.g. additive and multiplicative) on encrypted data. For instance, the Cheon, Kim, Kim and Song (CKKS) scheme \cite{cheon2017homomorphic} allows an unlimited number of both addition and multiplication operations to be run on real numbers, while the Brakerski, Gentry, and Vaikuntanathan (BGV) scheme \cite{dijk2010fully} also supports both operations but on integers.
            
        \end{itemize}
        
        \subsection{Homomorphic encryption classes}
        \paragraph{}
        Generally, the operations that can be performed on homomorphically encrypted data are classified into polynomial functions (e.g. addition, multiplication) and non-polynomial functions (e.g. division, square root). However, the schemes designed to allow polynomial functions to be performed may not allow non-polynomial functions to be efficiently performed, and vice versa. Therefore, researchers have classified HE schemes into two main classes:
        
        \begin{enumerate}
            \item Word-wise HE schemes refer to single-instruction-multiple-data (SIMD) style, where many plaintexts can be packed into one ciphertext. These kinds of schemes allow addition and multiplication operations to be efficiently performed on ciphertext. The CKKS \cite{cheon2017homomorphic}, BGV \cite{dijk2010fully}, and BFV \cite{fan2012somewhat} schemes are all classified as word-wise cryptosystems.
            
            \item Bit-wise HE schemes are efficiently able to implement operations such as division, max/min, and others using lookup tables (LUT). In these kinds of schemes, the plaintext gets encrypted bit by bit; consequently, it is much easier to present it as Boolean circuits. The FHEW \cite{ducas2015fhew} and TFHE \cite{chillotti2020tfhe} schemes are examples of bit-wise schemes.
        \end{enumerate}
        
        \subsection{Adopted platform}
        \paragraph{}
        As mentioned earlier, watermarking encrypted data using a CDW and nonlinear algorithm such as Sebe's algorithm \cite{sebe2005noise} requires performing word-wise and bit-wise operations on the encrypted data, and able to deal with real numbers. Therefore, the adopted encryption platform must be able to perform both kinds of operations. Indeed, Pegasus \cite{lu2021pegasus} is one of the most recent frameworks that can proficiently perform both; therefore, Pegasus was adopted in this research as an encryption platform.
        
        Pegasus can evaluate both polynomial and non-polynomial functions efficiently in terms of both time and memory consumption. It adopts a conversion algorithm to repack the FHEW ciphertext to CKKS ciphertext in sublinear time complexity instead of linear complexity. In fact, the key for the conversion algorithm is of size $\mathcal{O}(2\overline{n} \text{ } log \text{ } q)$, which makes it more effective compared to other frameworks.
        
        The Pegasus platform is built on the Simple Encrypted Arithmetic Library (SEAL) \cite{sealcrypto}. SEAL is an open-source encryption library developed by the Microsoft research team that supports BFV \cite{fan2012somewhat}, BGV \cite{dijk2010fully}, and CKKS \cite{cheon2017homomorphic} schemes. It is considered to be a levelled homomorphic encryption scheme, meaning the multiplicative depth of the cryptosystem must be predetermined. Hence, Pegasus requires the multiplicative depth to be predetermined. 

        The multiplicative depth represents the number of multiplications that can be performed on the encrypted data. As the multiplicative depth increases, the ciphertext size also increases. 
        
        
        \subsubsection{Pegasus functions}
        \paragraph{}
        In general, Pegasus's functionality focuses on the cloud side where it is assumed the data will always be encrypted. First, Pegasus encrypts the data in an RLWE ciphertext (CKKS) to perform polynomial functions efficiently. After that, if there is a need to perform non-polynomial functions, the framework will convert the RLWE ciphertext to an LWE ciphertext (FHEW); it uses a LUT to do this efficiently. Finally, Pegasus repacks the LWE ciphertext back into an RLWE ciphertext. The functionality of Pegasus can be summarised into four main functions: 
        
        \begin{enumerate}
        \item \textbf{The key-switching function ($\mathcal{F}_{KS}$)} is used to switch down the LWE ciphertext from a higher dimension to a lower dimension to improve the performance of the look-up table function (the next function). Note, this function introduces a negligible amount of error of around $2^{-14}$.
         \item \textbf{The look-up table function ($\mathcal{F}_{LUT}$)} can efficiently evaluate non-polynomial functions such as square root, division, and others on LWE ciphertext. It is designed to be able to process numbers of more than 40 bits long with a bounded error within the interval [$2^{-10},2^{-7}$]. This amount of error is acceptable in this research. Note that the message capacity that expected to be processed by this function must be predetermined; also, as the message capacity increases, the performance of the function decreases.
         
         \item \textbf{The linear transform function ($\mathcal{F}_{LT}$)} allows the LWE ciphertext to be repacked back to an RLWE ciphertext. Its approach is proven to be more efficient in terms of performance and memory consumption compared to the prior approaches.
         
         \item \textbf{The mod $q$ function ($\mathcal{F}_{mod}$)} is adopted to efficiently perform polynomial functions on the ciphertext. This function requires a multiplicative depth of nine in the original Pegasus implementation.
        \end{enumerate}

    \section{Data obfuscation}
    \subsection{Introduction}
        \paragraph{}
        Data obfuscation (DO) is an efficient technique to overcome the dilemma of sharing the data without breaching its privacy via reducing its sensitivity e.g. de-linking the values in each record to remove the connection between  \cite{Bakken2004}. The DO is an approach consisting of a set of techniques that can be adopted to lower data sensitivity in various ways while preserving its usefulness. Such an approach allows data to be shared with other users, such as scientists and researchers, without without revealing the values of individual data items \cite{Seidl2015}.
        
        DO is widely used in many applications including medical, scientific, business and others \cite{Bakken2004}. The reason behind its popularity is that it efficiently blurs the fine data to different degrees to suit the data owner's requirements  \cite{Panah2019}. The advantage of using DO is that the data remains usable after obfuscating, unlike encrypted data \cite{Hessler2012}. In other words, the content of data is unreadable after encryption, whereas after obfuscation it is still readable to a certain degree based on the DO technique used. 
        
        \subsection{Data obfuscation categories}
        \paragraph{}
        According to Bakken et al. \cite{Bakken2004}, DO techniques fall into three main categories: anonymisation, randomisation, and swapping techniques. The DO techniques are classified based on how they treat the data.
        
        \begin{itemize}
            \item \textbf{Anonymising techniques} anonymise each single data point by replacing it with a static or dynamic interval \cite{sweeney2002k}. In other words, each data point gets rounded and replaced by the nearest interval. The resulting values are not supposed to reveal any information about the original data points. For these kinds of techniques, the statistical results are maintained to a certain degree based on the interval size used to anonymise the data. The smaller the interval, the more accurate the results, while larger intervals give more anonymity to the data. 
            
            \item \textbf{Randomising techniques} create a copy of the dataset with different data points that returns accurate statistical results \cite{agrawal2000privacy}. These techniques work by dividing the data into subsets, then modifying the data points in each subset, and the resultant copy of the data will return statistical results as accurate as those run on the original dataset. These kinds of techniques prevent attackers from reconstructing a dataset by running the same queries repeatedly. 
            
            \item \textbf{Swapping techniques} work by changing the position of each data point in the same column \cite{gomatam2003distortion}. In other words, the actual value of each data point remains the same, only its position changes based on a specific mechanism. These kinds of techniques guarantee accurate results for many calculations including mean, median, max, variance and others. The main purpose of such a technique is to break the sequence of the data if it contains only one column, or if the dataset contains two columns or more, then it also de-links the values in each record. 
            
            
            
        \end{itemize}
        
        \subsection{Data obfuscation properties}
        \paragraph{}
        As there are many methods that can be categorised as DO techniques, researchers have summarised the general properties of DO techniques into three main types of properties: reversibility, specification, and shift properties \cite{Bakken2004}. The impact on those properties may vary depending on how the DO technique actually works.
        
        \begin{itemize}
            
            \item \textbf{The reversibility property} refers to the possibility that an adversary could deobfuscate an obfuscated dataset $X^{\prime\prime}$ given only $X^{\prime\prime}$. This property defines the security of the technique and will differ from one technique to another. The more complex it is to deobfuscate an obfuscated data $X^{\prime\prime}$ without the key $Obf_{key}$, the more secure the technique.
            
            \item \textbf{The specification property} indicates the parameter size for a particular DO technique. For instance, for anonymising techniques, it refers to the interval size, whereas for swapping techniques, it refers to how far two neighbouring points are from each other in the dataset. This parameter can be absolute or relative. An absolute parameter indicates that the amount of change will be the same for all datasets. In contrast, a relative parameter indicates that each dataset will be treated differently based on its magnitude (e.g. a certain percentage of its size). For example, the absolute feature in the swapping technique indicates that each element gets mapped to the same position for all datasets of the same size.
            
            \item \textbf{The shift property} points to the amount of distortion that can be added to the data after the obfuscation process. The amount of distortion differs based on the chosen technique. In other words, some techniques may apply distortion by constant values or random values. However, some DO techniques do not add any distortion to the data after the obfuscating process, for example, swapping techniques. Therefore, the latter kind of technique is a potential candidate to address the problem of reinforcing the security of the watermark verification process without affecting the data usability. 
            
        \end{itemize}
        
        \subsection{Adopted technique}
            \paragraph{}
            The main contribution of this research is to reinforce the security of the chosen watermarking algorithms in a way in which neither the embedded watermark nor the data usability for selected operations is affected. The security of the chosen watermarking algorithms will be reinforced in terms of verifying the embedded watermark using an efficient method. This step aims to prevent unauthorised users from verifying the watermark even if they have the watermark $K$ and the watermarked data $X^{\prime}$, which will nominate the cloud to become the neutral judge that can verify the data ownership. This section demonstrates the technique adopted to achieve this goal.  
            
            This research adopts data swapping as a DO technique to reinforce the security of the watermarking algorithms. This technique aims to break the original data sequence for each column separately, which will de-link the individual values in each row if the data has two columns or more. To illustrate, Table \ref{table:DOExample} provides an example of a simple dataset consisting of two columns representing the age and salary of five employees (rows) before and after obfuscation.
            
         \begin{table}[H]
        \caption{An example of the adopted DO technique}
        \label{table:DOExample}
                \centering
                \begin{subtable}{5cm}
                   \caption{\textbf{Before obfuscation}}
                   \label{table:DOExampleA}
                   \vskip0.2cm
                   \qquad
                  \begin{tabular}{|c|c|}
                   \hline
                   \textbf{Age} & \textbf{Salary} \\ \hline
                   39           & 6549            \\ \hline
                   26           & 9473            \\ \hline
                   52           & 11826           \\ \hline
                   32           & 8789            \\ \hline
                   22           & 7120            \\ \hline
                  \end{tabular}
                \end{subtable}
                 \begin{subtable}{5cm}
                  \caption{\textbf{After obfuscation}}
                  \label{table:DOExampleB}
                  \vskip0.2cm
                   \qquad
                 \begin{tabular}{|c|c|}
                  \hline
                  \textbf{Age} & \textbf{Salary} \\ \hline
                  26           & 8789            \\ \hline
                  22           & 11826           \\ \hline
                  39           & 7120            \\ \hline
                  52           & 6549            \\ \hline
                  32           & 9473            \\ \hline
                 \end{tabular}
                \end{subtable}
            \end{table}
            
            Table \ref{table:DOExampleA} shows the data before the obfuscation process and Table \ref{table:DOExampleB} shows the data after the obfuscation process. From these tables, it is clear that the technique does not change the actual values in the dataset; however, it changes the order of individual data points in each column separately. Consequently, the watermark will not be affected, nor will the data usability for selected operations. 
            
            This technique efficiently protects the watermark from being verified as follows. From Equation \ref{eqn:watermarkVerification} and the pseudocode from Algorithm \ref{VerifyingAlgo1}, it is clear that the watermark verification process for both algorithms depends on the sequence of the data. in Equation \ref{eqn:watermarkVerification}, each data point $\hat{x}_{i}$ in the watermarked dataset $\hat{X}$ has to match the corresponding element $s_{i}$ in the watermark sequence $S$ which is generated using the watermark $K$. Also, in Algorithm \ref{VerifyingAlgo1}, each data point $\hat{x}_{i}$ in the watermarked dataset $\hat{X}$ has to match the corresponding element in the original dataset $x_{i}$. That means if the data sequence in the dataset $\hat{X}$ is broken, then it will not be possible to verify the watermark even with a copy of the watermarked data and the watermark verification sequence. Taking advantage of that feature, the swapping technique will be quite efficient in preventing other users from verifying the data's ownership even if they own the watermark and the watermarked dataset under certain parameters. This feature is applicable even to data watermarked by Sebe's 2006 algorithm which will be used in future work.
            
            Gomatam and Karr \cite{gomatam2003distortion} covered a couple of the swapping techniques which work in the same way but are based on different mechanisms, such as the elementary swap and random swap techniques. The elementary swapping technique works by selecting two data points in the same column and then swapping them. The advantage of such a technique is to preserve the relationship between the swapped elements, which might be useful in some scenarios. The random swap technique manages to swap each data point independently within the same column. The advantage of this technique is that it is much harder to reverse engineer the obfuscated data $X^{\prime\prime}$ without the deobfuscation key $Obf_{key}$.
            
            The time complexity to find the correct order for a column $C$ consisting of $n$ nonidentical elements using a brute force attack is $\mathcal{O}(n!)$ which is the case in the watermarking algorithm 1. However, if the elements in the column $C$ are binary which is the case for the watermark verification in Sebe's algorithm, then the time complexity is $\mathcal{O}(\frac{n!}{r!(n-r)!})$ based on the combination formula ${ }_{n} C_{r}$ \cite{walpole1993probability}. That means this approach is secure in terms of the data reversibility under certain parameters as will be explained in the result and discussion chapter. Therefore, this research adopts the random swapping technique as a DO technique.

            This research developed an algorithm to obfuscate the data and another to deobfuscate based on a secret key $Obf_{key}$. The pseudocode of Algorithm \ref{ObfuscationAlgo} represents the obfuscation algorithm.
        
         \renewcommand{\thealgorithm}{3}
           
        \begin{algorithm}[H]
          \caption{Obfuscation Algorithm}\label{ObfuscationAlgo}
          \begin{algorithmic}[0]
          \State \textbf{Input:} The data $X^{\prime}[n]$ and the key $Obf_{key}$ are the inputs to the obfuscation function.
          \State \textbf{Output:} The obfuscated data $X^{\prime\prime}[n]$ is returned as an output.
        
          \begin{algorithmic}[1]
            \Function{Obfuscation}{}
            \For{i $\rightarrow$ 0 to $n$}
               \State  indexes[$Obf_{key}$[i]] $\gets$ i
          \EndFor
           \For{ i $\rightarrow$ 0 to $n$}
               \State   index $\gets$ indexes[i]
             \State swap $X^{\prime}$[i], $X^{\prime}$[$Obf_{key}$[i]]
          \State         $Obf_{key}$[index] $\gets$ $Obf_{key}$[i]
          \EndFor
            \EndFunction
            \end{algorithmic}
          \end{algorithmic}
        \end{algorithm}
        
        On the other hand, the pseudocode of Algorithm \ref{DeobfuscationAlgo} represents the mechanism to deobfuscate the data $X^{\prime\prime}[n]$ using the same secret $Obf_{key}$
        
        \renewcommand{\thealgorithm}{4}
        
        \begin{algorithm}[H]
          \caption{Deobfuscation Algorithm}\label{DeobfuscationAlgo}
        
          \begin{algorithmic}[0]
          \State \textbf{Input:} The obfuscated data $X^{\prime\prime}[n]$, and the secret key $Obf_{key}$ are the inputs to the deobfuscation algorithm
          \State \textbf{Output:} The watermarked data (deobfuscated) $X^{\prime}[n]$ is returned as an output.
        
          \begin{algorithmic}[1]
            \Function{deobfuscation}{}
            \For{i $\rightarrow$ 0 to $n$}
               \State  indexes[$Obf_{key}$[i]] $\gets$ i
          \EndFor
           \For{ i $\rightarrow$ 0 to n}
               \State   index $\gets$ indexes[i]
             \State  swap  $X^{\prime\prime}$[i], $X^{\prime\prime}$[indexes[i]]
          \State indexes[$Obf_{key}$[i]] $\gets$ indexes[i]
          \State $Obf_{key}$[index] $\gets$ $Obf_{key}$[i]
          \EndFor
            \EndFunction
            \end{algorithmic}
          \end{algorithmic}
        \end{algorithm}

    \section{Proposed framework}
        \paragraph{}
            
            The goal of the proposed framework is to allow the cloud to watermark encrypted data, in addition to giving the cloud control over the usability of the watermarked data. Another goal is to reinforce the watermark security via protecting the watermark verification process against unauthorised users even if they own the watermark value and the watermarked data, whereby only the cloud can verify it. In other words, the framework nominates the cloud to become the neutral judge to verify the watermark ownership and resolve ownership conflicts without affecting the data usability.
            
            To build such a framework, this research uses a combination of three privacy-preserving techniques: data watermarking (DW), homomorphic encryption (HE), and data obfuscation (DO). Each combination of these three techniques may lead to a different usage, which will have its own advantages and disadvantages. 
            
            Table \ref{table:UseCases} shows the potential combinations for the DW and the DO techniques on homomorphically encrypted data based on the chosen algorithms and techniques. In the table, a value of 1 indicates that the DO or DW technique was applied to the data first, a value of 2 indicates that the technique was applied to the data after the first technique, and a value of 0 indicates that the technique was not applied. 
            
            This research uses HE to secure data storage and transmission from the data owner (Alice) to the recipient (Bob) via the cloud. Thus, it is always assumed that the data is encrypted using Pegasus \cite{lu2021pegasus}; consequently, the HE stage has been excluded from the table. Note that this research relies on the original security parameters for Pegasus to secure the environment.
            
            From Table \ref{table:UseCases}, it is clear that each combination leads to a different usage which may be suitable for specific scenarios. The fourth combination was chosen for the proposed framework to achieve the purpose of this research.
            
            In the fourth combination, the encrypted data is first watermarked by either watermarking algorithm 1 or 2, and then, it is obfuscated using the highlighted obfuscation algorithm. This means that the obfuscation process will not only break the sequence of the data but also will break the sequence of the embedded watermark. Consequently, verifying the watermark will not be possible unless the data is first deobfuscated. In other words, only users who own the obfuscation key will be able to verify the watermark. This approach protects the watermark verification process without affecting the data usability, which achieves the goal of this research.
            
            The proposed framework assumes that the cloud is the party responsible for generating the sequence $S$ using the watermark $K$, which is known only to the cloud and the recipient. After that, the cloud embeds the sequence $S$ into the encrypted data $X$, which will result in watermarked data $X^{\prime}$. Then, the cloud obfuscates the watermarked data $X^{\prime}$ using the secret key $Obf_{key}$ which will result in obfuscated and watermarked data $X^{\prime\prime}$. The secret key $Obf_{key}$ must be kept by the cloud and not revealed to any other user. The watermarked copy of data created via this process will be unverifiable unless it is first deobfuscated using the same key $Obf_{key}$. In this scenario, the cloud has been nominated to be the only user who can verify the embedded watermark.
            
\begin{figure}[hp!]
                \begin{center}
                    \includegraphics[width=17.5cm,height=16cm,] {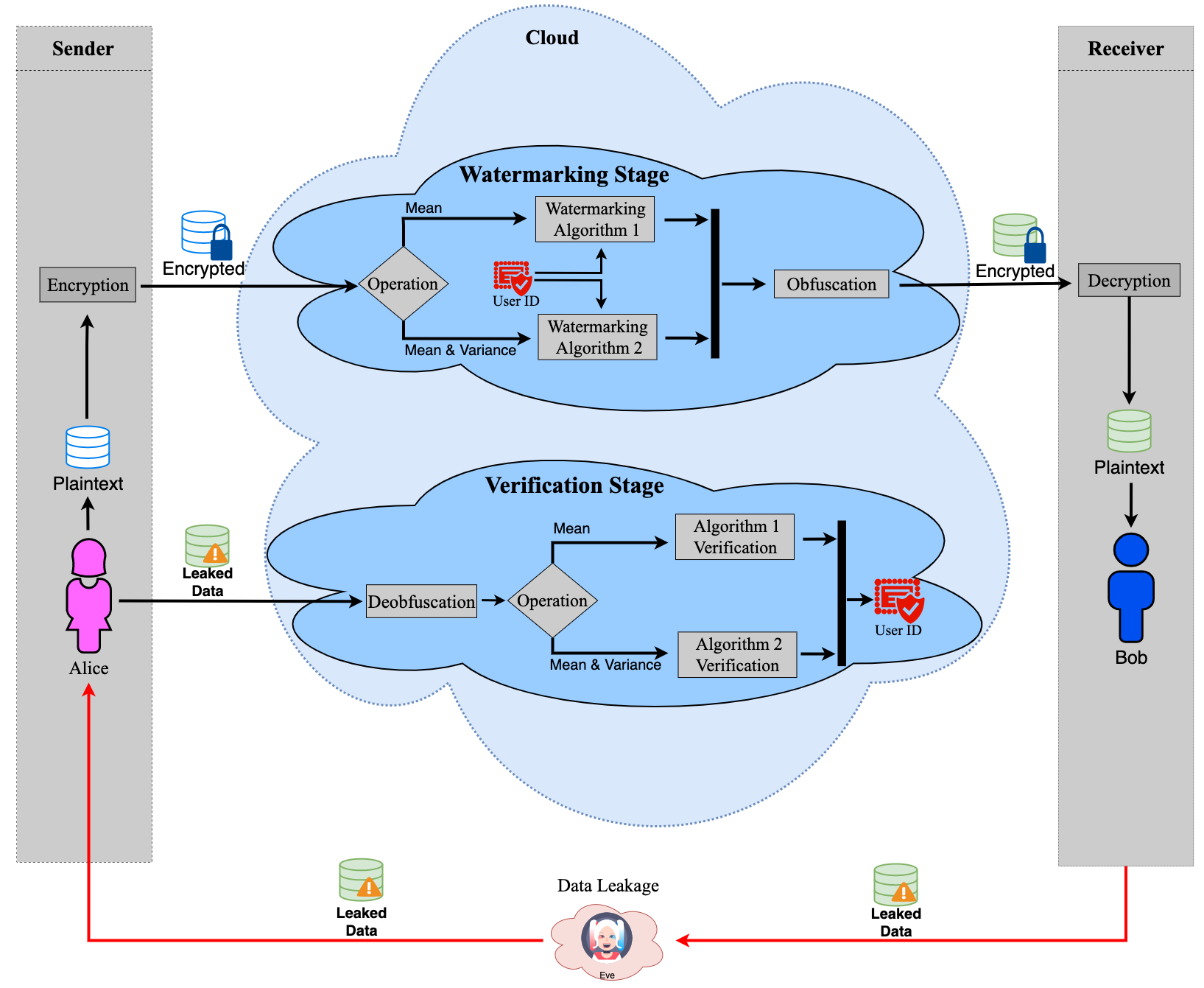}
                    \caption{The proposed framework}
                    \label{proposedFramework}
                \end{center}
            \end{figure}

            To illustrate the framework, Figure \ref{proposedFramework} highlights the scenario and the role for each party. The scenario is that Alice decides to send a copy of the data to Bob; however, she wants to ensure that if Bob leaks the data, then he can be identified. Also, Bob wants to know the unique and secret value that his data has been watermarked with.


            First, Alice encrypts the data using Bob's public key. Then, she sends the encrypted data $X$ to the cloud with the type of statistics that she want to be preserved, e.g. the mean only or the mean and variance. Next, the cloud receives the encrypted data $X$, and uses the most suitable algorithm to embed the watermark $K$, using Bob's public credentials (e.g. $Bob_{RK}$, $Bob_{SwK}$, and others). For example, the cloud will use watermarking algorithm 1 to preserve the mean only, or algorithm 2 to preserve the mean and variance. After that, the cloud obfuscates the watermarked and encrypted data $X^{\prime}$ using the secret key $Obf_{key}$ known only to the cloud. Finally, the cloud sends the data $X^{\prime\prime}$ to Bob, who can decrypt it using his private key.
            
            In this case, even though Bob knows the watermark $K$, he cannot detect it because the data is obfuscated. In order to detect the watermark, the data must be deobfuscated first.
            
            Assuming that Bob has leaked the data, first the cloud must deobfuscate the data using the secret key $Obf_{key}$. Then, the cloud can verify the data ownership using the correct verification algorithm. 
            
            In conclusion, the proposed framework has contributed to finding a balance between the recipient's right to know the secret value that has been embedded into their copy of data \cite{frattolillo2006web}, while at the same time, it protects the watermark security by preventing the recipient from verifying it \cite{cox2007digital}. 
            
            Table \ref{table:keys&parameters} shows the keys and parameters used in the proposed framework and specifies whether they are public keys or owned by specific users.
            
            \begin{table}[H]
            \caption{Essential keys and parameters used in the proposed framework}
            \label{table:keys&parameters}
            \centering
                \begin{tabular}{|c|c|c|}
                \hline
                    \textbf{Name} & \textbf{Type} & \textbf{Known to} \\ \hline
                    \hline
                    $Bob_{pk}$                  & key           & Public                        \\ \hline
                    $Bob_{SwK}$                 & key           & Public                        \\ \hline
                    $Bob_{EK}$                  & key           & Public                        \\ \hline
                    $Bob_{RK}$                  & key           & Public                        \\ \hline
                    $Bob_{RotK}$                & key           & Public                        \\ \hline
                    $Bob_{RelK}$                & key           & Public                        \\ \hline
                    $Bob_{s}$                   & key           & Bob                           \\ \hline
                    $Obf_{key}$                 & key           & Cloud                         \\ \hline
                    $q$                         & parameter     & Public                        \\ \hline
                    $\textbf{g}_{digit}$        & parameter     & Public                        \\ \hline
                    $\textbf{g}_{rns}$          & parameter     & Public                        \\ \hline
                    $\Delta$                    & parameter     & Public                        \\ \hline
                    $\overline{n}$              & parameter     & Public                        \\ \hline
                    $n^\prime$                  & parameter     & Public                        \\ \hline
                    $\underline{n}$             & parameter     & Public                        \\ \hline
                    $K$                         & parameter     & Cloud and Recipient (Bob)     \\ \hline
                    $M$                         & parameter     & Cloud                         \\ \hline
                    $\epsilon$                  & parameter     & Cloud                         \\ \hline
                    
                \end{tabular}
            \end{table}


            \begin{table}[H]
            \caption{Use cases for each combination using obfuscation, watermarking, or both}
        \label{table:UseCases}
\begin{tabular}{|c|cc|p{8.1cm}|}
\hline
{\textbf{No.}} & \multicolumn{2}{c|}{\textbf{Operations}}                          & \multicolumn{1}{c|}{\textbf{Details}}     \\ \cline{2-3}                        & \multicolumn{1}{c|}{\textbf{Obfuscation}} & \textbf{Watermarking} & \multicolumn{1}{c|}{}\\ 
\hline
1                            & \multicolumn{1}{c|}{1}                    & 0                     & \begin{tabular}[c]{@{}l@{}}* Data sequence is broken.\\  * The ownership of the data is not verifiable.  \\  * Use case: it could be useful in cases where\\rank order operations (max, medium) are\\required along with other statistical operations\\(mean, variance).\end{tabular}                                                           \\ \hline
2                            & \multicolumn{1}{c|}{0}                    & 1                     & \begin{tabular}[c]{@{}l@{}}* The data sequence is preserved.\\  * The individual data elements are blurred in a\\way in which the accuracy of the results\\ of statistical operations is not affected.\\  * The ownership of the data is verifiable.\\ * The watermark verification security is not\\protected, whereby any user who knows the\\ watermark will be able to verify it.
\\ * Use case: it could be useful in cases where\\individual data elements are watermarked to\\identify the ownership, while the data's\\sequence is also significant, e.g. ECG signals.\end{tabular}                                                                        \\ \hline
3                            & \multicolumn{1}{c|}{1}                    & 2                     & \begin{tabular}[c]{@{}l@{}}* The data sequence is broken.\\  * The individual data elements are blurred in\\a way in which the accuracy of the results\\ of statistical operations is not affected.\\  * The ownership of the data is verifiable. \\ * The watermark verification security is not\\protected. 
\\ * Use case: it could be useful when the data's\\ownership is identifiable, and the sequence of\\the data is insignificant. 
\end{tabular}                                                                                               \\ \hline
4                            & \multicolumn{1}{c|}{2}                    & 1                     & \begin{tabular}[c]{@{}l@{}}* The data sequence is broken.\\  * The individual data points are watermarked.\\  * The watermark is protected by a second layer\\of protection, whereby the data needs to be\\deobfuscated first to verify its ownership.\\  * Use case: it is useful when data's ownership\\needs to be identifiable, and its sequence is\\insignificant. For example, the data owner wants\\data analysts to analyse the data using\\statistical calculations such as sum, mean,\\variance, and others, to get a better\\understanding of the data.\\Moreover, in this case, the user who wants to\\verify the ownership of the watermark must\\have the obfuscation key to first deobuscate\\the data before verifying its ownership.\end{tabular} \\ \hline
\end{tabular}
\end{table}

\chapter{Experiments}
\paragraph{}
    This chapter highlights the requirements to conduct the experiment to ensure the results can be reproduced. The first section explains the dataset used in this research. The second section describes the experimental environment and the last section discusses the settings used in the experiments.  

    \section{The dataset}
    \paragraph{}
    To verify the functionality of the proposed framework, this thesis uses the "Smart Home Dataset with Weather Information" \cite{datasetreference}. The data was collected from a smart house containing multiple IoT devices. The IoT devices include home appliances, weather sensors, and others. The data contains readings from those devices over 350 days at a rate of one span per minute in kilowatts (KW). 
    
    The dataset consists of 32 columns and 50,3910 observations stored in a comma-separated values (CSV) file. Of the 32 columns, 27 columns contain decimal values, three columns contain strings, and two contain integers.
    
    As this research focuses on univariate data, the experiment was conducted using one column, which is the "House overall [kw]" column. This column contains decimal values, which makes it suitable to be watermarked using an algorithm such as Sebe's algorithm \cite{sebe2005noise}. 
    
    A feature of this dataset is that its values are relatively small. For example, the mean of the values in the chosen column is roughly 0.86 and the variance is approximately 1.12.
    
    \section{Experimental environment}
    \paragraph{}
    The experiment was conducted on a MacBook Pro laptop running the macOS Big Sur operating system. The processor was a 2.8 GHz Intel core i7, and, the computer had 16 GB of memory.
    
    \section{The experimental settings}
    \paragraph{}
    As mentioned earlier, this research assumes that the data will be watermarked and obfuscated after it has been encrypted using Pegasus \cite{lu2021pegasus} which is written in C++ programming language. Therefore, all the watermarking and obfuscation algorithms are coded using the same language.
    
    In this experiment, the research uses the default settings used in the original implementation of Pegasus, except for the multiplicative depth. The multiplicative depth represents the number of operations that can be performed on a ciphertext, and as the multiplicative depth for a ciphertext increases, the ciphertext size increases. In the original implementation of Pegasus, the multiplicative depth was set to 13 after enabling the repacking feature. However, this experiment only required the multiplicative depth to be 12, which reduces the ciphertext size and improves the performance.
    
    Table \ref{table:experimentSettings} highlights the settings used in this research.
    
    \begin{table}[H]
    \caption{The experiment settings}
    \label{table:experimentSettings}
        \centering
        \begin{tabular}{|c|c|}
            \hline
            \textbf{Parameter} & \textbf{value}         \\ \hline
            $Q$                       & 735             \\ \hline
            $\overline{n}$            & $2^{16}$        \\ \hline
            $n^\prime$                & $2^{12}$        \\ \hline
            $\underline{n}$           & $2^{10}$        \\ \hline
            $\Delta$                  & $2^{40}$        \\ \hline
            $B_{ks}$                  & $2^{7}$         \\ \hline
            $h$                       & 64              \\ \hline
            $msg$                     & [-8 , 8]        \\ \hline
            S2C multiplier            & 1.0             \\ \hline
            moduli                    & 12              \\ \hline
            $\epsilon$                & 0.05            \\ \hline
            $K$                       & 0xB61B          \\ \hline
        \end{tabular}
    \end{table}

\chapter{Results and discussion}
    \paragraph{}
    The aim of this research was to build a watermarking framework that will allow a third party (the cloud) to watermark encrypted data for ownership protection applications. The framework will also nominate the cloud as the only user able to verify the embedded watermark. The contribution of this research is to strengthen the security of the watermarking algorithms against passive attacks using an efficient method and without affecting the data usability for selected operations. This chapter presents and interprets the thesis findings based on these objectives.
    
    The first section of this chapter analyses and presents the thesis results. The second section interprets and discusses those results. 
    
    \section{Results analysis}
    \paragraph{}
    Results collected from the experiments include the accuracy of the results of selected operations conducted on the watermarked data; the size of the data after encryption; and performance in regard to encrypting the data, generating and embedding the watermark in the data, obfuscating the data, and decrypting the data based on the settings defined in the experiments chapter. The results of each of these are presented in a separate section.
    
        \subsection{Accuracy of the results of statistical operations}
        \paragraph{}
        Processing homomorphically encrypted data introduces noise that may interfere with the usefulness of the watermarked data after decryption \cite{zheng2012walsh}. To verify the amount of noise, this study watermarked the same data using the same parameters in both encrypted and unencrypted domains. After that, the difference between the results of the selected statistical operations run on each copy of the data was calculated. The following table shows the differences between data watermarked in an encrypted and an unencrypted domain for each watermarking algorithm.
        
        \begin{table}[H]
        \caption{Differences between the accuracy of the results of operations run on data watermarked in encrypted and unencrypted domains using watermarking algorithms 1 and 2}
        \label{table:ResultAccuracy}
        \centering
            \begin{tabular}{|l|c|c|}
                \hline
                \multicolumn{1}{|c|}{\textbf{Operation}} & \textbf{Algorithm 1} & \textbf{Algorithm 2}                  \\ \hline
                sum                         & $-3.3 \text{ x}10^{-08}$  & $-2.54\text{ x}10^{-07}$                      \\ \hline
                mean                        & $-3.2\text{ x}10^{-08}$   & $-4.13\text{ x}10^{-06}$                      \\ \hline
                variance                    & N/A                       & $-2.58\text{ x}10^{-05}$                      \\ \hline
                standard deviation          & N/A                       & $-1.74\text{ x}10^{-05}$                      \\ \hline
            \end{tabular}
        \end{table}
    
        \subsection{Memory consumption}
        \paragraph{}
        This section highlights the memory required to store and process the ciphertext based on the stated parameters in the experiments chapter. The experiments found that the size of ciphertext objects was always consistent. The size of a single ciphertext object was always roughly 11.53 MB, while the ciphertext size for the complete encrypted data was always around 92.3 MB.
        
        \subsection{Performance analysis}
        \paragraph{}
        This section analyses the performance of encrypting the data, generating the watermark, embedding it, decrypting the data, and finally obfuscating the data. The performance is measured based on processing datasets of two sizes: the complete dataset which contains 503,910 data points, and the block format dataset where each block contains only 65,536 data points in order to get a better insight into how the performance may differ based on dataset size. Furthermore, to eliminate the random delay that could occur in individual experiments, each experiment was executed ten times and the results were averaged. 
        
\begin{figure}[H]
        \begin{center}
            \includegraphics[width = 6in, height = 5in,]{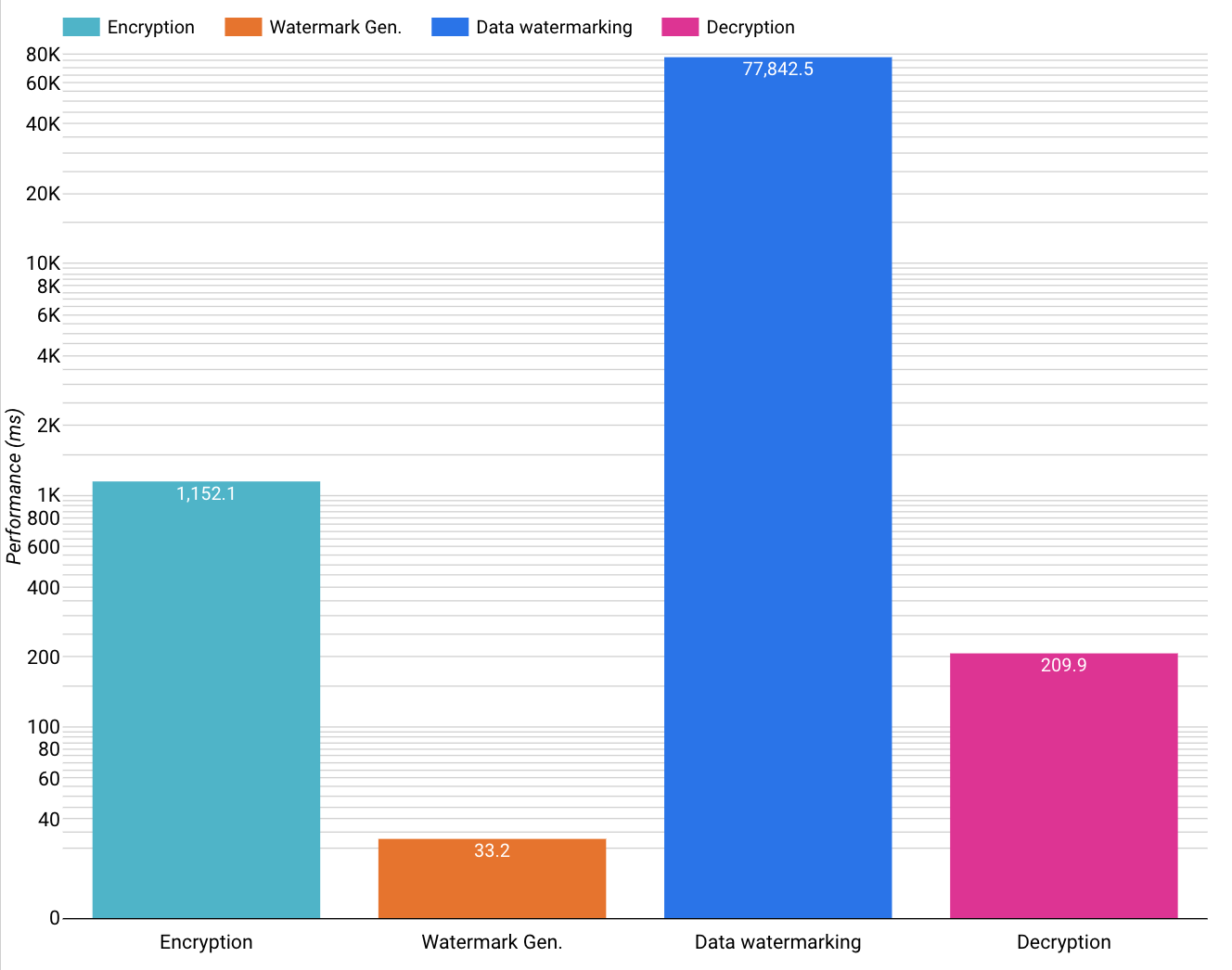}
            \caption{Performance of watermarking algorithm 1 on the complete dataset}
            \label{Algorithm1Averages}
        \end{center}
        \end{figure}

     Figure \ref{Algorithm1Averages} shows the average time required to encrypt, generate the watermark, embed, and decrypt the dataset in milliseconds (ms) using watermarking algorithm 1 on the complete dataset.
        
         The graph shows that watermarking the data was the most time-consuming process, taking around 77,842.5 ms, whereas the watermark generation process was the least time-consuming at about 33.2 ms. It also shows that the time required to encrypt the data was roughly 1152.1 ms, while the decryption process required only 209.9 ms.
        
\begin{figure}[H]
        \begin{center}
            \includegraphics[width = 6in, height = 5in,]{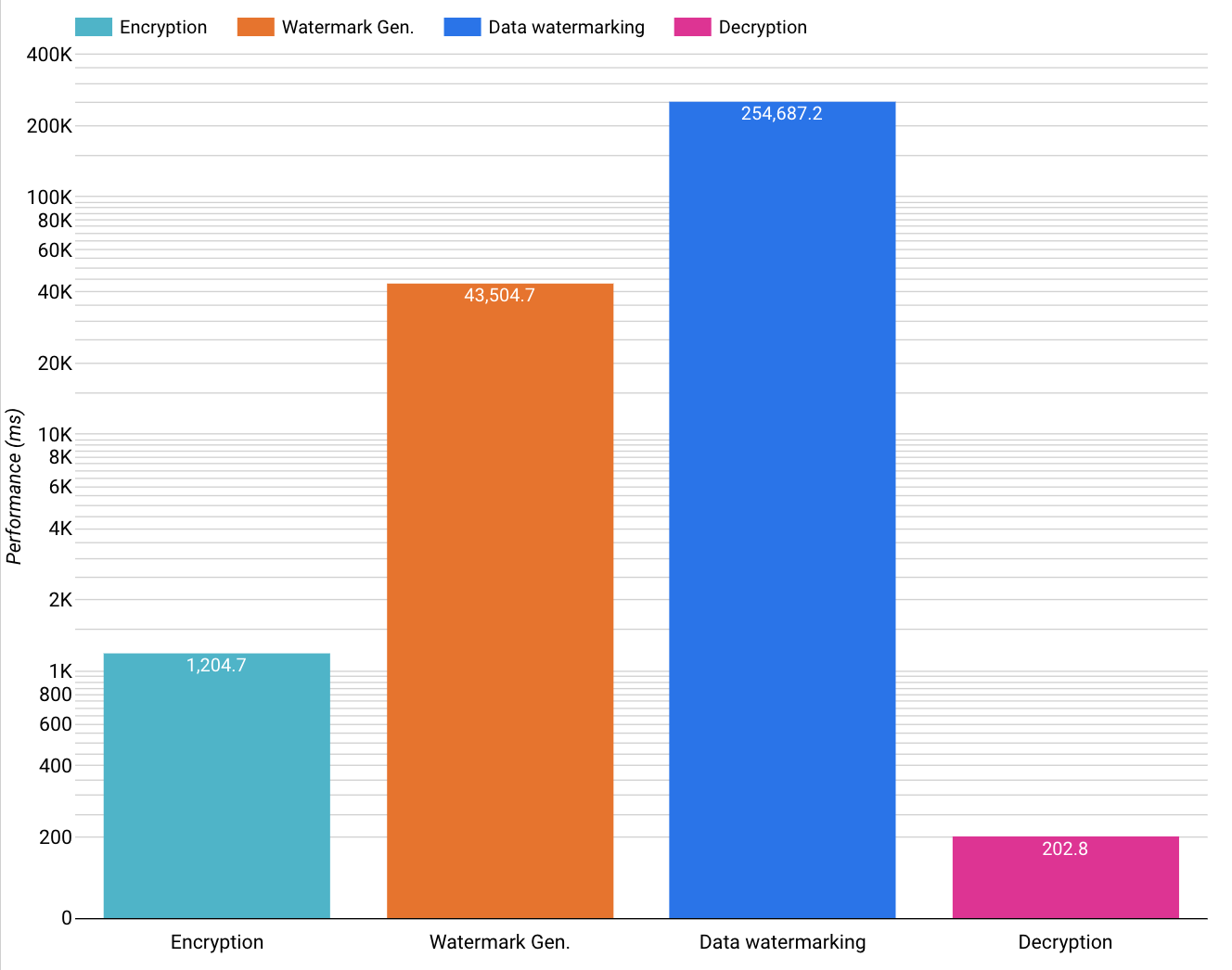}
            \caption{Performance of watermarking algorithm 2 on the complete dataset}
            \label{Algorithm2Averages}
        \end{center}
        \end{figure}

        Figure \ref{Algorithm2Averages} shows the average time required to encrypt, generate the watermark, embed, and decrypt the dataset in ms using watermarking algorithm 2 on the complete dataset.
        
        The graph shows that watermarking the data was the most time-consuming process at around 254,687.2 ms, whereas the decryption process was the least time-consuming process at about 202.8 ms. Also, it shows that the watermark generation process took about 43,504.7 ms, while the time required to encrypt the data was roughly 1204.7 ms.
        
\begin{figure}[H]
        \begin{center}
            \includegraphics[width = 6in, height = 5in,]{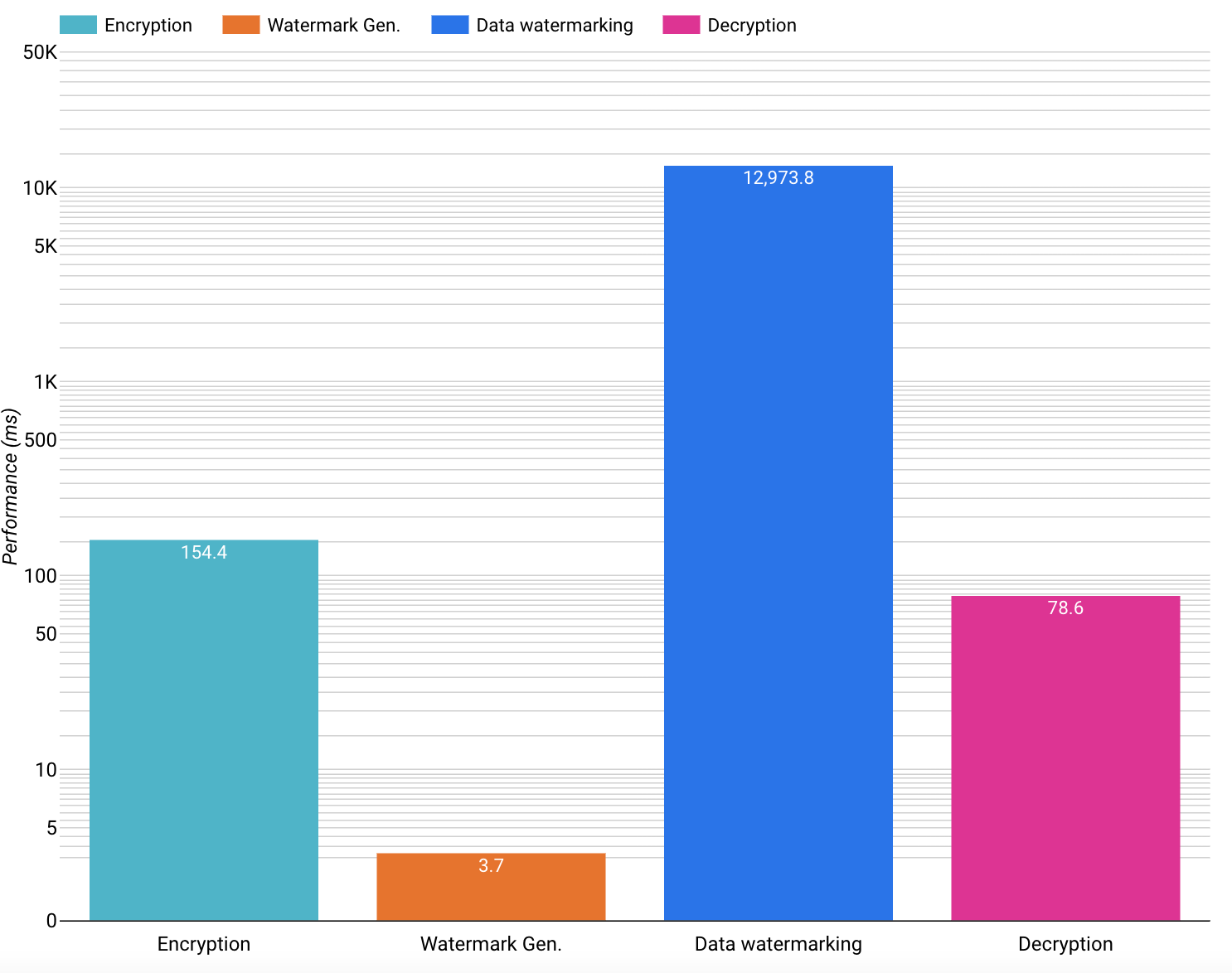}
            \caption{Performance of watermarking algorithm 1 on the blocked dataset}
            \label{Algorithm1Averages-BlockedData}
        \end{center}
        \end{figure}

        Figure \ref{Algorithm1Averages-BlockedData} shows the average time required to encrypt, generate the watermark, embed it, and decrypt data in ms using watermarking algorithm 1 on the blocked dataset.
        
        The graph shows that watermarking the data was the most time-consuming process at around 12973.8 ms, whereas watermark generation was the least time-consuming process at only about 3.7 ms. It also shows that the decryption process took about 78.6 ms, while the data encryption took roughly 154.4 ms.
        
\begin{figure}[H]
        \begin{center}
            \includegraphics[width = 6in, height = 5in,]{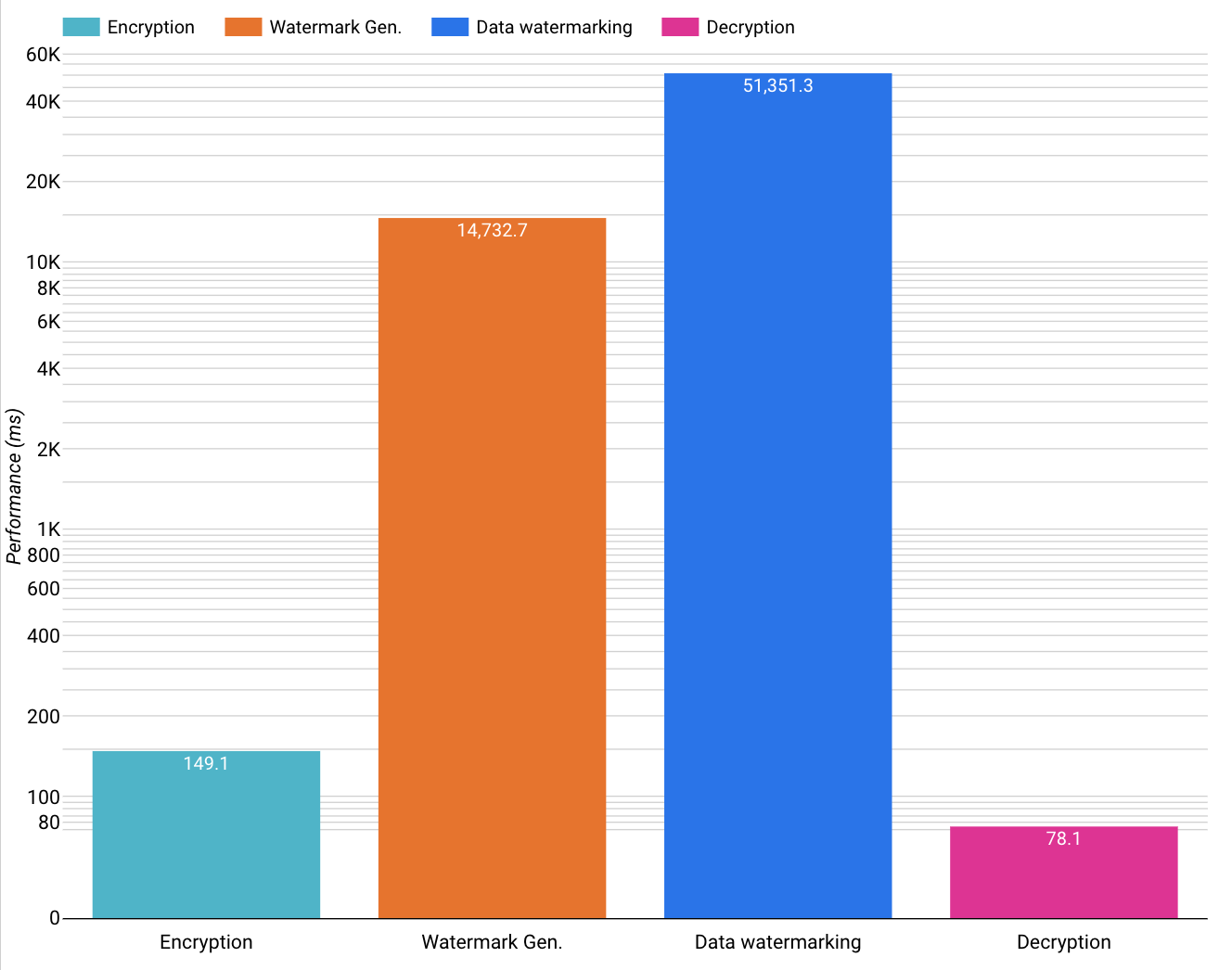}
            \caption{Performance of watermarking algorithm 2 on the blocked dataset}
            \label{Algorithm2Averages-BlockedData}
        \end{center}
        \end{figure}

        Figure \ref{Algorithm2Averages-BlockedData} shows the average time required to encrypt, generate the watermark, embed it, and decrypt in ms using watermarking algorithm 2 on the blocked dataset.
        
        The graph shows that watermarking the data was the most time-consuming process at around 51,351.3 ms, whereas the decryption was the least time-consuming process at about 78.1 ms. It also shows that the watermarking generation process took about 14,732.7 ms, while the data encryption took roughly 149.1 ms.

\begin{figure}[H]
            \centering
            \begin{minipage}{0.45\textwidth}
                \centering
                \includegraphics[width=0.9\textwidth]{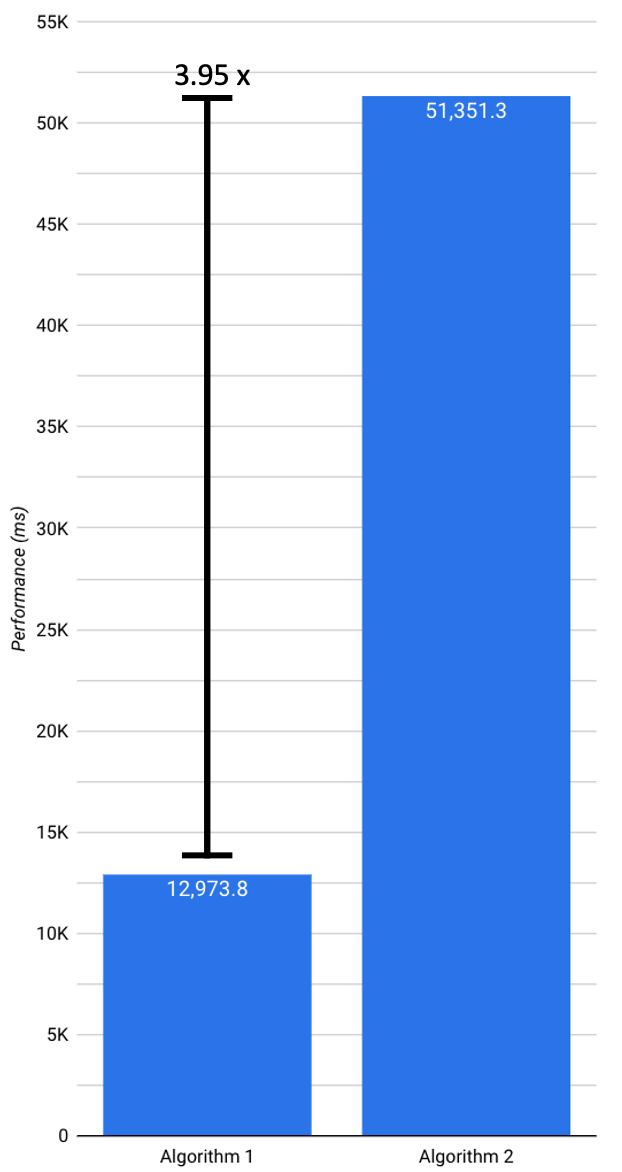} 
                \caption{Performance of watermarking algorithms 1 and 2 on the blocked data}
                \label{DifferenceBetweenBothAlgoBlocked}
            \end{minipage}\hfill
            \begin{minipage}{0.45\textwidth}
                \centering
                \includegraphics[width=0.9\textwidth]{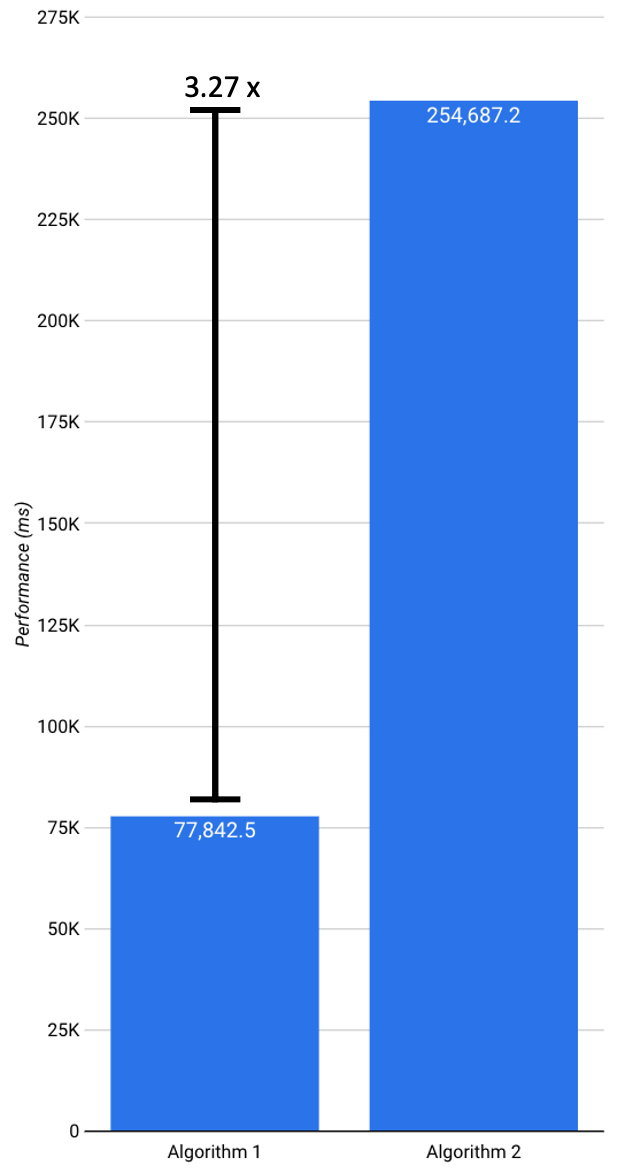} 
                \caption{Performance of watermarking algorithms 1 and 2 on the complete data}
                \label{DifferenceBetweenBothAlgoCompleted}
            \end{minipage}
        \end{figure}

        Figure \ref{DifferenceBetweenBothAlgoBlocked} (on the blocked data) and \ref{DifferenceBetweenBothAlgoCompleted} (on the complete data) compare the time consumed by the watermarking process for each algorithm.
        
        The two figures show that watermarking algorithm 1 was faster than the watermarking algorithm 2 on both dataset sizes. From Figure \ref{DifferenceBetweenBothAlgoBlocked}, it is clear that watermarking 65,536 data points using algorithm 1 took roughly 12,973.8 ms, while watermarking the same number of data points using algorithm 2 consumed around 51,351.3 ms, which means watermarking algorithm 1 was 3.95 faster than watermarking algorithm 2 for the watermarking process. It can be see in Figure \ref{DifferenceBetweenBothAlgoCompleted} that watermarking 50,3910 data points using algorithm 1 required 77,842.5 ms, whereas watermarking algorithm 2 needed roughly 254,687.2 ms, making watermarking algorithm 1  3.27 times faster than algorithm 2 for this process.
        
        Refer to the appendix for results regarding time consumed for individual experiments. 
        
\begin{figure}[H]
        \begin{center}
            \includegraphics[width = 6in, height = 4in,]{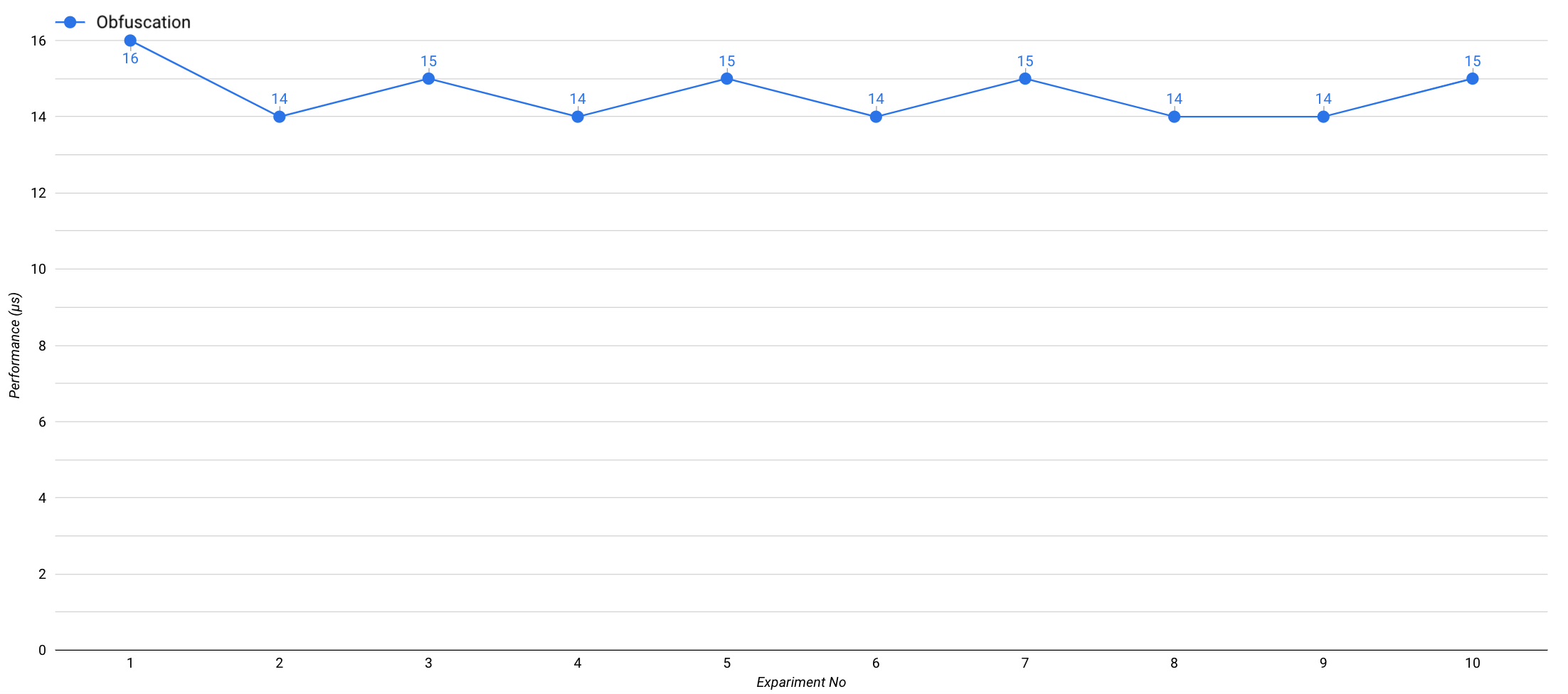}
            \caption{Performance (time in $\mu$s) for obfuscating data over 10 experiments}
            \label{ObfuscatingFunctionRunningTime}
        \end{center}
         \end{figure}

        Figure \ref{ObfuscatingFunctionRunningTime} shows the time required to obfuscate the encrypted data in microseconds ($\mu$s) for the 10 experiments.
        
        The graph shows that the highest value was 16 $\mu$s, while the lowest was 14 $\mu$s. On average, the obfuscating process required roughly 14.6 $\mu$s to obfuscate the encrypted data. It is worth mentioning that a similar amount of time was consumed to obfuscate each dataset.
        
    
    \section{Results discussion}
    \paragraph{}
    In this section, the results presented in the previous section are discussed and interpreted.
    
    As stated earlier, performing arithmetic operations on encrypted data introduces extra noise to the data \cite{zheng2011implementation}. Thus, some researchers have recommended using a specific type of watermarking algorithm that does not require heavy computations over the ciphertext, such as the WHT algorithm \cite{zheng2012walsh}. However, this thesis showed that state-of-the-art encryption platforms such as Pegasus \cite{lu2021pegasus} are able to watermark encrypted data using watermarking algorithms which require heavy computations including multiplication, square root, and division operations, such as Sebe's algorithm \cite{sebe2005noise} with a very negligible amount of noise. 
    
    Table \ref{table:ResultAccuracy} shows that the differences between the results of selected operations run on data watermarked while encrypted and those results from data that was watermarked as plaintext were extremely negligible, that is, roughly from $-3.3 \text{ x}10^{-08}$ to $-1.74\text{ x}10^{-05}$ using the parameters that were explained in the experiments chapter. It is also noticeable that watermarking algorithm 1 introduced less noise compared to watermarking algorithm 2 for both the sum and average calculations. The reason for this is that algorithm 2 required more heavy computations including the square root and multiplication operations on the encrypted data. The amount of noise introduced by both algorithms is very acceptable in this research.
    
    In terms of memory consumption, the size of a single ciphertext was around 11.53 MB based on the settings in the experiments chapter, whereas from the original Pegasus implementation \cite{lu2021pegasus}, the size of a single ciphertext object was about 12.58 MB. Hence, the proposed framework saves more than one MB per ciphertext. In general, the ciphertext size is affected by two major factors: the multiplicative depth of a ciphertext, and its packing capacity. In the proposed framework, every single ciphertext had the ability to handle up to 11 multiplicative operations (i.e. a multiplicative depth of 12). Also, each ciphertext can pack up to $2^{16}$ encrypted data points. In other words, each ciphertext was able to store 65,536 encrypted elements; consequently, the whole dataset could be packed into only eight ciphertext objects, with a total size of about 92.3 MB. Note that the ciphertext remained a fixed size, that is, the size before and after processing remained the same.
    
    In terms of performance, this study used the levelled HE to construct the proposed framework, which is slower than the partial HE used in the majority of previous studies \cite{xiang2018database}, \cite{zhang2021hope}. However, the partial HE is not sufficient to watermark data using an algorithm such as Sebe's algorithm in an encrypted domain because Sebe's algorithm requires a variety of arithmetic operations to be performed on the ciphertexts, as explained in the methodology chapter. In this implementation, the performance was mainly affected by the message capacity, the multiplicative depth, and the ciphertext size.
    
    From Figures \ref{Algorithm1Averages} and \ref{Algorithm2Averages}, and \ref{Algorithm1Averages-BlockedData} and \ref{Algorithm2Averages-BlockedData}, it is clear that the time consumed by the encryption and decryption processes in both algorithms was almost the same for each dataset. This is because the size of the datasets were the same for each dataset before and after the watermarking process for both algorithms. 
    
    In addition, the graphs show that the watermark generation process using algorithm 1 was significantly faster compared to that using algorithm 2 even though the time complexity for both is $\mathcal{O}(n)$. The reason for this is that the watermark generation process in algorithm 1 did not need to process the encrypted data because it is not a CDW algorithm. Whereas the watermark generation process in algorithm 2 was the second most time-consuming process because of the homomorphical computations needed to be performed on the encrypted data. As algorithm 2 is a CDW algorithm, the watermark generation process must be implemented for each block of data, whereas in algorithm 1, the watermark generation process needs to be performed only once and can be used for all blocks of data. 
    
    Furthermore, Figure \ref{Algorithm2Averages-BlockedData} shows that the watermark generation process was 3.48 times faster than the data watermarking process on the blocked data, taking into consideration that it needs to be done for each block of data. On the other hand, Figure \ref{Algorithm2Averages} shows that the watermark generation process was 5.85 times faster than the data watermarking process, considering that it needs to be done only once for the complete data. Therefore, it is highly recommended to watermark data as a complete dataset, rather than in block format, or at least in larger blocks using a CDW algorithm such as Sebe's algorithm \cite{sebe2005noise}. 
    
    In addition, it was clear that the watermarking process was the most time-consuming process for both algorithms and on both dataset sizes. Although Algorithm \ref{WatermarkingAlgo1} and Equation \ref{eqn:watermarking} show that the watermarking process in both algorithms is of complexity $\mathcal{O}(n)$, Figures \ref{DifferenceBetweenBothAlgoBlocked} and \ref{DifferenceBetweenBothAlgoCompleted} show that in algorithm 1 it was faster by more than three times compared to algorithm 2. The reason is that algorithm 2 needs to perform the multiplication and addition multiple times on the encrypted data, while algorithm 1 requires addition only. The experiment showed that multiplications operations are slower than additions.
    
    Figure \ref{ObfuscatingFunctionRunningTime} highlights the additional performance overhead for the obfuscation function. Note that the time was calculated in $\mu$s for the obfuscation process. In contrast, in Figures \ref{Algorithm1Averages}, \ref{Algorithm2Averages}, \ref{Algorithm1Averages-BlockedData}, and \ref{Algorithm2Averages-BlockedData} the time was shown in ms for other functions. Hence, the obfuscation function was remarkably fast compared to all other functions in this study. In fact, it required on average only 14.6 $\mu$s to obfuscate the dataset and therefore it is considered a lightweight process. 
    
    In terms of the watermark verification security, in the case of watermarking algorithm 1, the probability of finding all possible permutations for a list of $n$ elements is $n!$ possible combination which is a significant number of combinations as $n$ gets larger; however, watermarking algorithm 1 was used for illustrative purposes only, and it is not secure against potential watermarking attacks as illustrated in the methodology section. On the other hand, calculating all possible combinations for a binary sequence, as is the case for watermarking algorithm 2 \cite{sebe2005noise}, will result in $(\frac{n!}{r!(n-r)!})$ possible combinations, if the adversary knows the correct value of M (which he or she is not supposed to) as mentioned in the methodology chapter. Therefore, if watermarking algorithm 2 watermarks a dataset of 150 data points, this will leave any adversary who tries to access the data with roughly $9.2\text{ x}10^{43}$ possible combinations, which is more than the combinations for the AES-128 \cite{daemen2001reijndael} which results in around $3.4\text{ x}10^{38}$ possible combinations. Thus, obfuscating the data using the proposed method will secure the watermark against unauthorised users. Note, this is also applied to the Sebe's 2006 algorithm \cite{sebe2006watermarking} which will be used in future work.
    
    \chapter{Conclusion}
    \paragraph{}
    This study was conducted with two main objectives. The first objective was to determine the circumstances that can allow watermarking of numerical data for ownership verification purposes by a third party in an encrypted domain. For the sake of this objective, it was essential to determine the amount of distortion introduced to data watermarked in an encrypted domain compared to data watermarked in an unencrypted domain. In addition, the study investigated the amount of memory required to store and process the encrypted data. Furthermore, it sought to discover how efficient it is to watermark encrypted data using selected existing algorithms. Another main objective of this research was to increase the robustness of the security of the chosen watermarking algorithms in a way whereby it becomes significantly difficult for an adversary to verify the embedded watermark. Based on these objectives, this chapter explores the main conclusions and limitations of this study, and proposed future works, each in a separate section.
    
    \section{Conclusion}
        \paragraph{}
        Indeed, this thesis contributed by successfully building a watermarking framework which has several advantages. The first advantage is that the framework allows a third party to watermark univariate and encrypted data for ownership protection applications using irreversible watermarking algorithms while controlling the usability of the watermarked data. The second advantage is that the framework creates a balance between the recipient's right to know the watermark that was inserted in her or his copy of the data and to protect the watermark security by protecting the watermark from being verified by the recipient.
        
        The research showed that it is possible to watermark encrypted data with a very insignificant amount of error using state-of-the-art encryption platforms. The difference between watermarking encrypted and unencrypted data was found to be roughly between $-3.3 \text{ x}10^{-08}$ and $-1.74\text{ x}10^{-05}$. In addition, it was found that watermarking algorithm 2 introduced more noise than watermarking algorithm 1 because of the intensive calculations needed to be performed on the encrypted data. Therefore, it was clear that as the number of calculations increases, the noise also increases. 
        
        Furthermore, the study examined the memory required to store and process the ciphertext via the framework and the factors that may affect it. Based on the settings specified in the experiments chapter, the size of each ciphertext was 11.53 MB and each was able to pack up to 65,536 encrypted data points. The experiment also clarified that the ciphertext size was not affected by the watermarking processes of either algorithm. In fact the ciphertext remained a fixed size, before and after the watermarking processes, for both algorithms.
        
        The performance of watermarking algorithms 1 and 2 was analysed to highlight their advantages and disadvantages in terms of performance. Watermarking algorithm 2 was approximately three times slower than watermarking algorithm 1, because algorithm 2 needed to perform more computations on the encrypted data than algorithm 1. 
        
        In addition, the research highlighted that the obfuscation technique suggested to reinforce the security of the chosen watermarking algorithms demonstrated remarkably fast performance compared to the encrypting, watermark generation, dataset watermarking, and decrypting processes. In other words, obfuscating the data added a negligible burden to the overall performance of the framework.
        
        
        Moreover, the proposed technique is considered secure against brute force attacks, as the time complexity for an adversary to explore all possible combinations is $\mathcal{O}(\frac{n!}{r!(n-r)!})$ for a binary watermarking sequence. In other words, for a binary sequence of 150 elements, an adversary would need to explore around $9.2 \text{ x} 10^{43}$ possible combinations, which is more than the possible combinations for AES-128.
            
        \section{Limitations}
        \paragraph{}
        There are two main limitations in this research. The first limitation is that watermarking algorithm 1 is not secure against potential attacks as explained in the methodology chapter. In fact, this algorithm was used for illustrative purposes only. The second limitation is that the chosen encryption platform requires the message capacity to be predetermined. In other words, the cloud would need to know the range of numbers that might be evaluated using the LUTs function, which will leak some information about the encrypted data.
        
        \section{Future Work}
        \paragraph{}
        Further research is required to develop a secure irreversible watermarking algorithm to be used for ownership protection applications, which is able to preserve the mean only. The algorithm must also be robust against attacks that might be launched against ownership protection watermarks including noise and forgery attacks. It is preferable that the algorithm does not require many computations on the actual data, so that it will be more efficient in terms of performance on encrypted data.
        
        In addition, further research might explore other encryption platforms which could be more efficient and not require the message capacity for LUTs function to be predetermined. This feature would improve the encryption environment in the framework in terms of maintaining the secrecy of the processed data.  
        
        Furthermore, an essential progression to advance this framework would be to extend the statistical calculations that can be preserved on the encrypted data to include the covariance of bivariate data. This step will generalise the framework and allow it to preserve more statistical calculations and also give it the ability to process bivariate datasets.

\appendix
\chapter{Additional results}

    \begin{figure}[H]
        \begin{center}
            \includegraphics[width = 6in, height = 3.5in,]{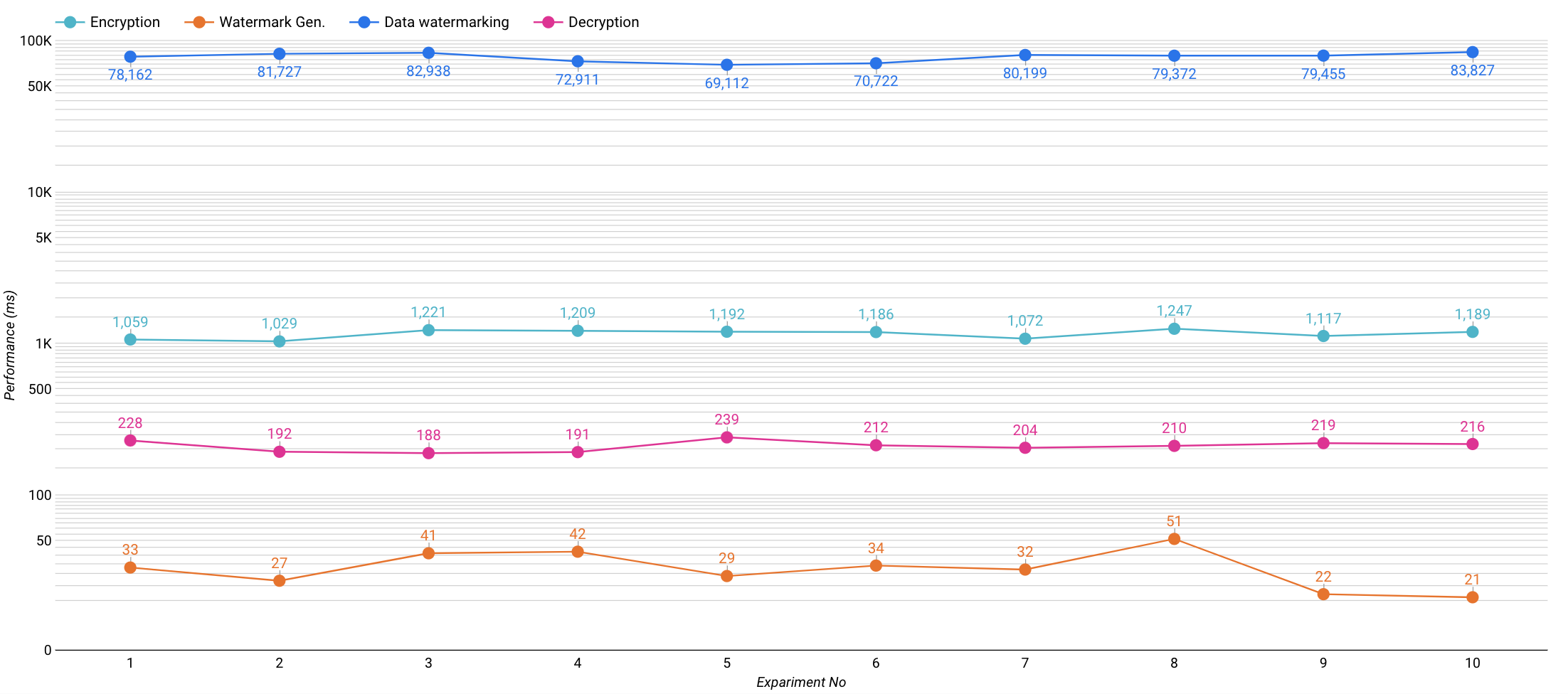}
            \caption{Performance of watermarking algorithm 1 on the complete dataset for all experiments}
        \end{center}
    \end{figure}

\begin{figure}[H]
        \begin{center}
            \includegraphics[width = 6in, height = 3.5in,]{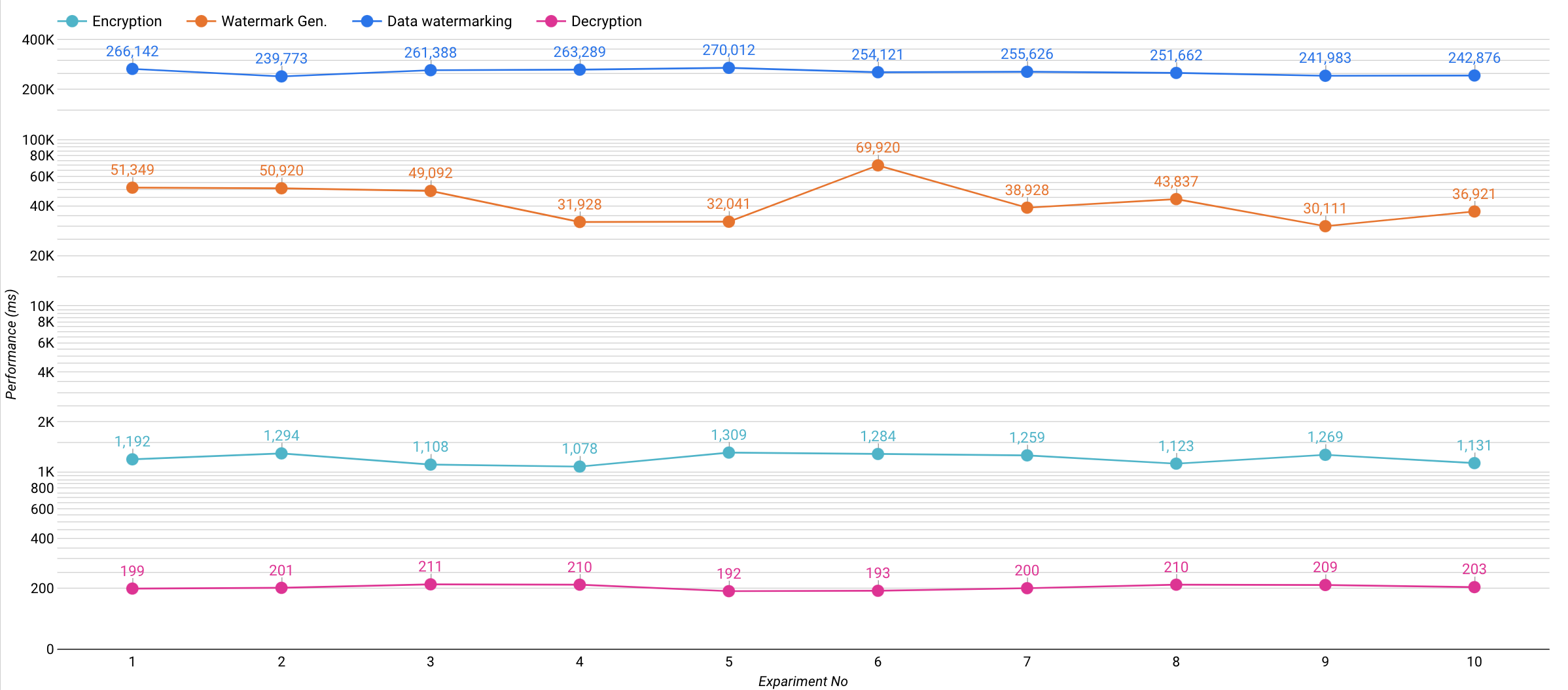}
            \caption{Performance of watermarking algorithm 2 on the complete dataset for all experiments}
        \end{center}
    \end{figure}
    
    \begin{figure}[H]
        \begin{center}
            \includegraphics[width = 6in, height = 3.5in,]{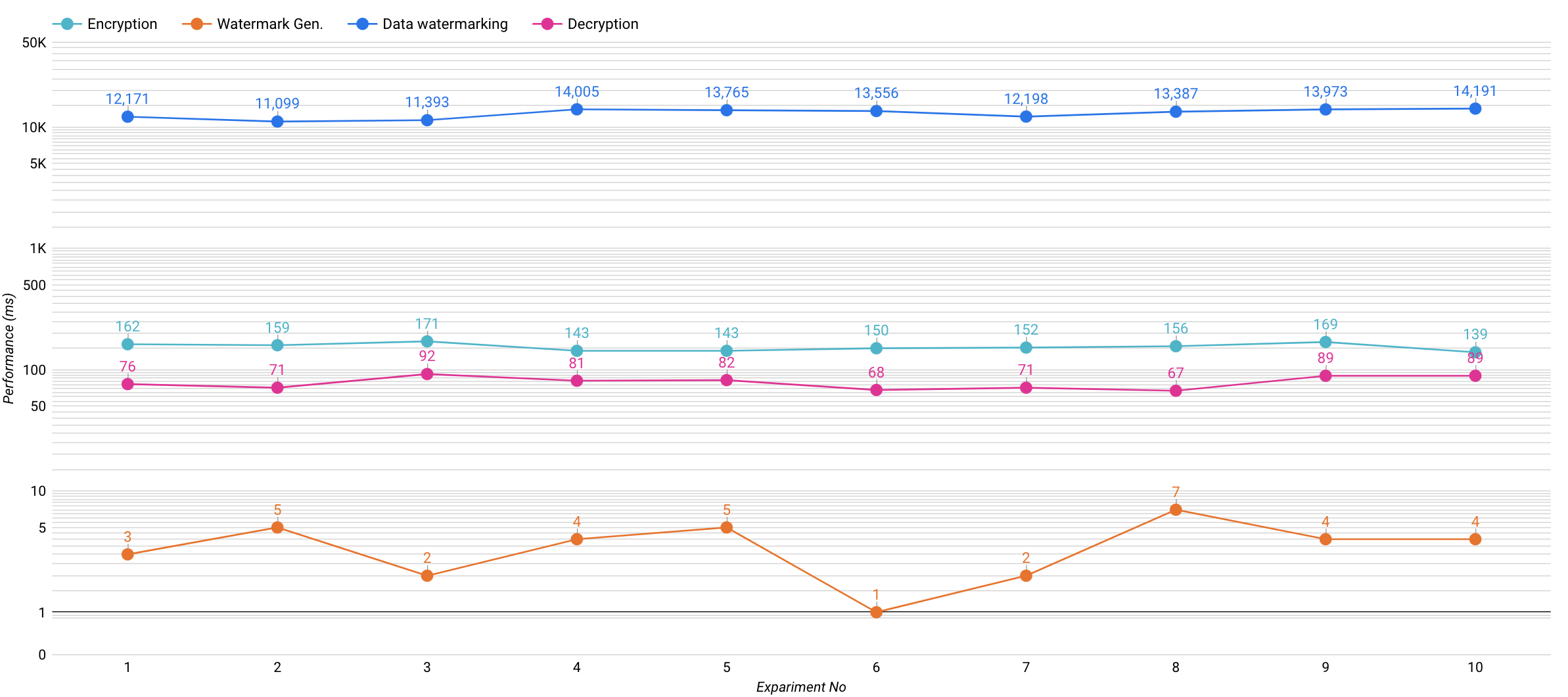}
            \caption{Performance of watermarking algorithm 1 on the blocked dataset for all experiments}
        \end{center}
    \end{figure}
    
    \begin{figure}[H]
        \begin{center}
            \includegraphics[width = 6in, height = 3.5in,]{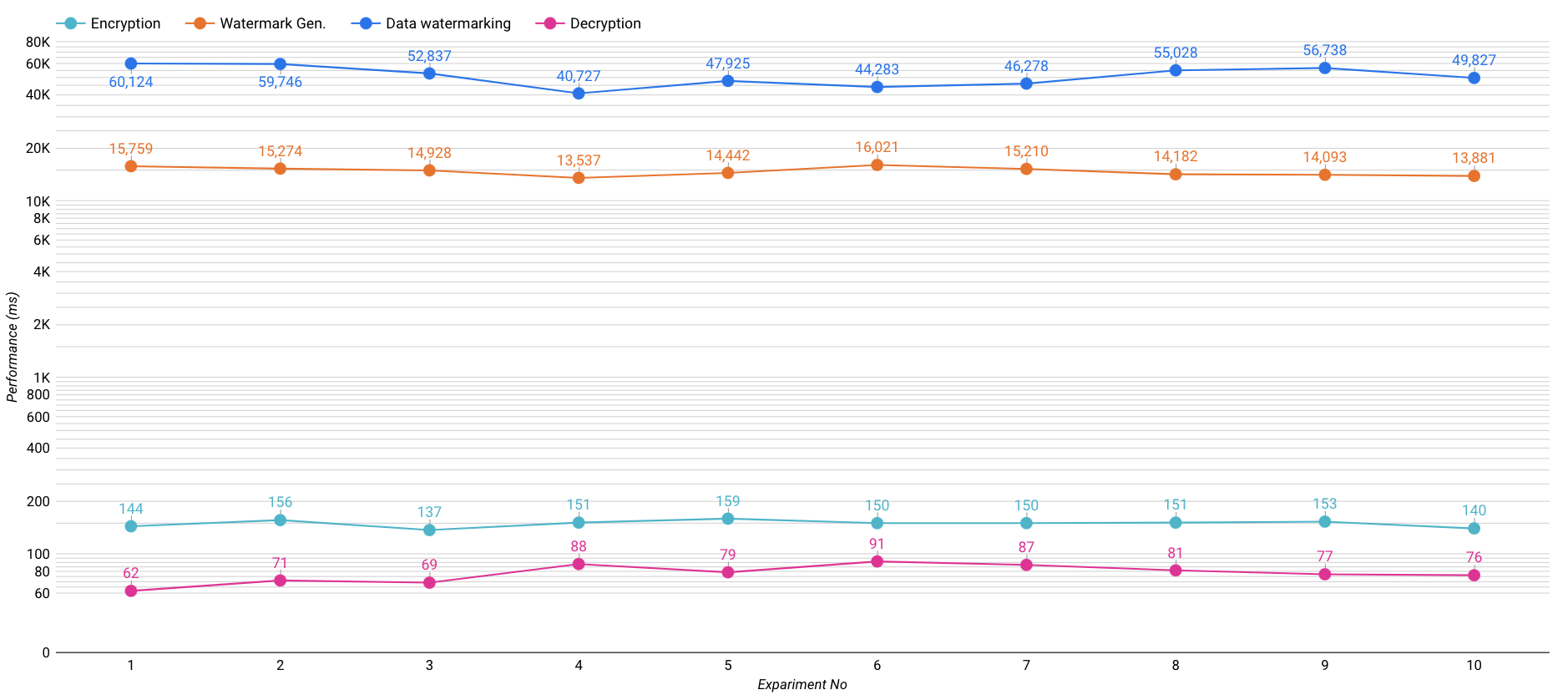}
            \caption{Performance of watermarking algorithm 2 on the blocked dataset for all experiments}
        \end{center}
    \end{figure}
    
\bibliographystyle{IEEEtran}
\bibliography{main}

\end{document}